%% file: arxiv-main.tex
\documentclass[sigconf, balance=false]{acmart}
\AtBeginDocument{%
  }

\setcopyright{acmlicensed}
\acmISBN{978-1-4503-XXXX-X/2018/06}
\usepackage{hyperref}

\usepackage{algorithm}
\usepackage[noend]{algpseudocode}
\usepackage[utf8]{inputenc} % allow utf-8 input
\usepackage[T1]{fontenc}    % use 8-bit T1 fonts
\usepackage{hyperref}       % hyperlinks
\usepackage{url}            % simple URL typesetting
\usepackage{booktabs}       % professional-quality tables
\usepackage{amsfonts}       % blackboard math symbols
\usepackage{microtype}      % microtypography
\usepackage{xcolor}         % colors

\usepackage{amsmath} 
\usepackage{multirow}
\usepackage{tabularx}
\usepackage{subfigure}
\usepackage{booktabs}
\usepackage{graphicx} 
\usepackage{colortbl}
\usepackage{makecell}
\usepackage{arydshln}
\usepackage{wrapfig}
\usepackage{caption}
\usepackage{subcaption}
\usepackage{xcolor}
\usepackage[most]{tcolorbox}
\def\method{MemShot}

\begin{document}

\title{Memory Shot for Long-Term Dialogue}

\author{Chunyi Peng}
\authornote{ \ \ indicates equal contribution.}
\affiliation{%
  \institution{Northeastern University}
  \city{Shenyang}
  \country{China}
}
\email{pengchunyi@mails.neu.edu.cn}

\author{Haidong Xin}
\authornotemark[1]
\affiliation{%
  \institution{Northeastern University}
  \city{Shenyang}
  \country{China}
}
\email{xinhaidong@stumail.neu.edu.cn}

\author{Xuanshuo Sheng}
\affiliation{%
  \institution{Northeastern University}
  \city{Shenyang}
  \country{China}
}
\email{shengxuanshuo@gmail.com}

\author{Xin Dai}
\affiliation{%
  \institution{Northeastern University}
  \city{Shenyang}
  \country{China}
}
\email{daix1@mails.neu.edu.cn}

\author{Zhenghao Liu}
\authornote{ \ \ indicates corresponding author.}
\affiliation{%
  \institution{Northeastern University}
  \city{Shenyang}
  \country{China}}
\email{liuzhenghao@mail.neu.edu.cn}

\author{Shuo Wang}
\affiliation{%
  \institution{Tsinghua University}
  \city{Beijing}
  \country{China}
}
\email{wangshuo.thu@gmail.com}

\author{Yukun Yan}
% \authornotemark[2]
\affiliation{%
  \institution{Tsinghua University}
  \city{Beijing}
  \country{China}
}
\email{yanyk.thu@gmail.com}

\author{Zulong Chen}
\affiliation{%
  \institution{Alibaba Group}
  \city{Hangzhou}
  \country{China}}
\email{zulong.czl@alibaba-inc.com}

\author{Yu Gu}
\affiliation{%
  \institution{Northeastern University}
  \city{Shenyang}
  \country{China}}
\email{guyu@mail.neu.edu.cn}

\author{Ge Yu}
\affiliation{%
  \institution{Northeastern University}
  \city{Shenyang}
  \country{China}}
\email{yuge@mail.neu.edu.cn}

\renewcommand{\shortauthors}{Chunyi Peng et al.}

\input{sections/abstract}




\maketitle

\input{sections/intro}
\input{sections/related_work}

\input{sections/formulation}

\input{sections/experimental_methodology}

\input{sections/experiments}

\input{sections/conclusion}
\bibliographystyle{ACM-Reference-Format}
\bibliography{sample-base}

\clearpage
\newpage
\appendix
\input{sections/appendix}
\end{document}

%% file: sections/abstract.tex
\begin{abstract}
Large Language Models (LLMs) have demonstrated strong capabilities in general conversation, instruction following, and complex reasoning. 
However, in long-term dialogue settings, they often struggle to locate and utilize historical information that is most relevant to the current query. 
Existing approaches attempt to address this issue by employing sophisticated memory construction methods, which maintain structured text-centered memory units through compressing and reorganizing user interaction history for memory maintenance and updating. 
However, these memory systems often rely on brute-force extraction of crucial evidence to associate episodes across different dialogue sessions, resulting in substantial computational overhead and weakening structural cues in the original interactions, such as speaker transitions, turn boundaries, and local contextual relationships. 
To avoid fragile text-based memory representations, we propose \textbf{\method{}} to redefine memory construction by leveraging dialogue structuring for long-term dialogue modeling and relying on the model's internal visual reasoning capabilities to associate key episodes within dialogues. 
Specifically, \method{} directly renders local contiguous dialogue spans into structured visual memory units, explicitly preserving meta-information and the chronological structure of dialogue turns while avoiding heavy-weight textual memory construction. 
Experimental results show that \method{} achieves stable and competitive performance on both LoCoMo and LongMemEval, while substantially shortening the memory construction pipeline and delivering \textbf{70$\times$} speedup. 
Further analysis reveals that \method{} enhances both the localization and utilization of historical evidence, while directing the model’s memory processing toward structured local dialogue cues and away from surface-level lexical matching in a flat text stream.
All codes are released on \url{https://github.com/NEUIR/MemShot}.
\end{abstract}

%% file: sections/intro.tex
\section{Introduction}
Large Language Models (LLMs) have demonstrated strong capabilities in general conversation, instruction following, and complex reasoning~\cite{seed2025seed2, yang2025qwen3, zeng2026glm}. However, due to the lost-in-the-middle problem~\cite{liu2024lost}, LLMs often struggle to extract salient cues that are truly relevant to the current query as dialogue histories grow longer. Consequently, establishing long-range dependencies among semantically related content becomes increasingly challenging for LLMs~\cite{liu2025comprehensive, wang2024beyond}. To mitigate these issues, existing approaches typically adopt retrieval-augmented methods~\cite{liu2026knowledge, zhou2025llm, lewis2020retrieval}, enabling LLMs to access relevant knowledge from the previous user interactions. Specifically, these methods segment user interaction histories into local chunks and organize them into retrievable units, thereby facilitating the selection of query-relevant content to support response generation. While such strategies partially alleviate the burden of modeling the entire dialogue history, they remain insufficient for maintaining an evolving state to deal with endless user interactions and organizing fine-grained, structurally coherent, and contextually related clues embedded within dialogues.
\input{figures/latency_acc}

To overcome these limitations, recent work~\cite{hu2026evermemos, fang2025lightmem, kang2025memory} has further advanced memory systems for managing long-term dialogue histories by transforming retrievable chunks into more manageable memory units. These units are typically represented as textual chunks that aggregate key information from dialogue segments sharing the same topic or episode. Building on this paradigm, subsequent methods~\cite{li2025memos, xin2026metamem, mao2025meta} continuously incorporate essential knowledge from dialogue history through iterative processes such as updating, compression, integration, and retrieval. As shown in Figure~\ref{fig:latency}, although these memory systems effectively capture salient information across extended interactions, they rely on increasingly heavy memory construction pipelines that repeatedly rewrite and reorganize consistent information from the original dialogue. This process leads to higher memory construction latency, especially when larger models are employed or when smaller LLMs require more inference steps. In contrast, a more fundamental yet underexplored question remains: do we truly require such heavyweight memory construction mechanisms, or are we effectively using a sledgehammer to crack a nut?

To answer this question, we draw further inspiration from how human long-term memory is organized and retrieved. From a cognitive science perspective, unlike existing memory construction methods, long-term memory is not naturally structured as a flat stream of textual fragments. Instead, continuous experience can be segmented into memory units through scene construction and higher-level event organization~\cite{hassabis2007deconstructing, zeidman2015constructing}, which in turn facilitates information encoding and retrieval~\cite{nolden2024prediction, laing2025event}. Motivated by these insights, we introduce \textbf{\method{}}, a novel framework that constructs memory units by converting each dialogue session into a structured ``memory shot''. Specifically, \method{} directly renders each local dialogue span into a lightweight, retrievable visual memory unit, while explicitly preserving its meta-information and chronological turn structure. This design fully leverages the advantages of visual encoding to keep each memory unit structurally coherent and self-contained, while avoiding the overhead of heavyweight memory construction and effectively exploiting the visual understanding capabilities of MLLMs for memory modeling.

Experimental results demonstrate the effectiveness and efficiency of \method{}, achieving competitive performance with existing memory-augmented generation models while delivering 70$\times$ faster memory construction. Leveraging the structurally rich visual memory, \method{} enables more precise retrieval of query-relevant historical evidence and supports more effective memory utilization, allowing the MLLM to capture crucial information from user interactions for response generation. Further analysis reveals that \method{} shifts the memory-augmented generation to a shooting-replay mechanism, which moves away from summarizing compact but fragile memory units~\cite{xin2026metamem} and instead leverages more comprehensive dialogue information through visual structures. This encourages capturing more relevant information from the dialogue and reduces dependence on surface-level lexical matching in a flat text stream, offering a more promising paradigm for long-form dialogue modeling.

%% file: figures/latency_acc.tex
\begin{figure}[t]
\includegraphics[width=0.95\linewidth]{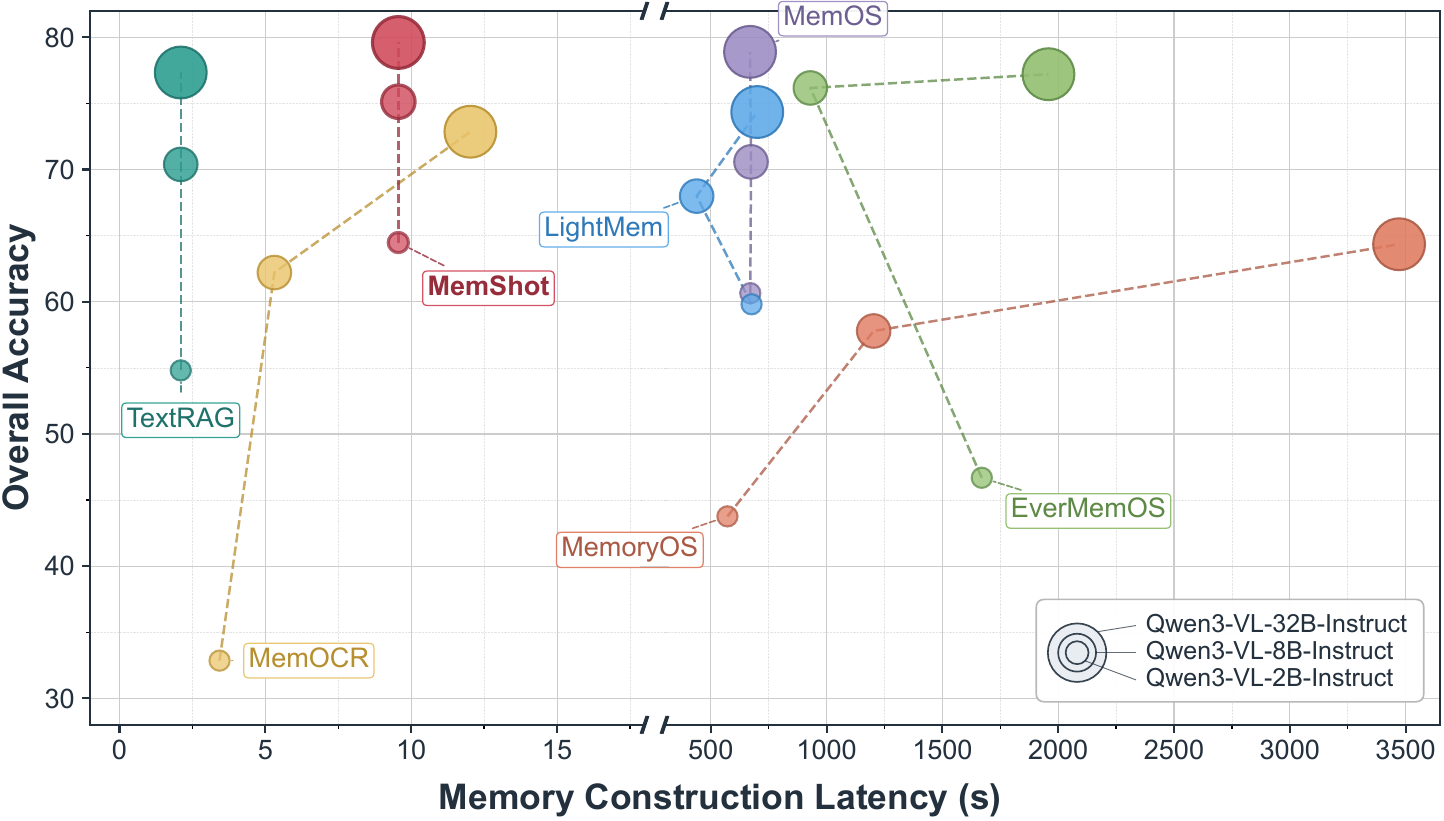}
\caption{Visualization of Memory Construction Latency and Answer Accuracy on the LoCoMo Dataset. We report results across different Qwen3-VL model scales.}
\label{fig:latency}
\end{figure}

%% file: sections/related_work.tex
\section{Related Work}
Large Language Models (LLMs) have demonstrated strong capabilities across a wide range of complex tasks~\cite{seed2025seed2, yang2025qwen3, zeng2026glm}. However, maintaining coherent long-term dialogue remains challenging, as relevant information becomes increasingly difficult to capture from extended interactions~\cite{liu2025comprehensive, wang2024beyond}, due to the ``lost-in-the-middle'' problem~\cite{liu2024lost}. To address this issue, a line of work introduces Retrieval-Augmented Generation (RAG) methods~\cite{liu2026knowledge, zhou2025llm, lewis2020retrieval} that segment long dialogue histories into text chunks and select query-relevant content for downstream response generation. Nonetheless, these RAG systems still face difficulties in capturing clues from long-form dialogues in prior experiments, especially when operating in dynamic and open-ended environments~\cite{wu2025sgmem,tan2025prospect,hu2026evermemos}.

To alleviate these issues, recent studies have further developed memory systems to maintain a consistent state for long-horizon user interaction modeling, continuously initializing and updating memory units by assimilating clues from long-form user interactions.
Earlier approaches like MemoryBank~\cite{zhong2024memorybank} and MemGPT~\cite{packer2023memgpt} focus on explicit memory storage and management beyond the immediate context window.
Currently, existing methods focus on different aspects to effectively construct memory units~\cite{du2025rethinking,hu2025memory,jiang2026anatomy}. 
For example, to reduce the noise in user interactions, LightMem~\cite{fang2025lightmem} proposes a hierarchical compression method that discards unnecessary text from the dialogue to accelerate memory construction. 
Mem0~\cite{chhikara2025mem0} and Zep~\cite{rasmussen2025zep} study salient memory extraction and graph-based organization, while A-MEM~\cite{xu2025mem} focuses on memory evolution and update mechanisms. 
MemU\footnote{Open-source memory infrastructure: \url{https://github.com/NevaMind-AI/memU}} organizes memories as a file-system-like hierarchy to support proactive long-running agents.
MemOS~\cite{li2025memos} and MemoryOS~\cite{kang2025memory} treat memory as an operating system, emphasizing construction through operations for memory management. 
To enhance the dependency among fragmented memory units, some methods like EverMemOS~\cite{hu2026evermemos} focus on better integrating and reorganizing memory units, helping to form coherent and stable knowledge structures that support long-horizon reasoning of LLMs.
Despite their differences in implementation, these methods largely rely on maintaining more effective memory units by retrieving related knowledge from long-form dialogues and applying updating, compression, integration, and retrieval operations to maintain and refine textual memory over time. 
Even when effective, these constructed memory units may weaken structural cues such as turn boundaries, speaker transitions, and local relations between neighboring utterances.

Thriving on the visual understanding capabilities of multi-modal LLMs (MLLMs)~\cite{bai2025qwen3, wang2025internvl3, yu2025minicpm}, recent work has begun to revisit long-context modeling by exploring the compression of long textual contexts into images through rendering.
DeepSeek-OCR~\cite{wei2025deepseek} and Glyph~\cite{cheng2025glyph} demonstrate that long text can be mapped into more compact visual representations, enabling more information to be preserved under limited context budgets~\cite{wang2024leveraging, lu2024text, li2025text}.
Furthermore, MemOCR~\cite{shi2026memocr} extends these text compression advantages to memory settings by maintaining structured rich-text memory and rendering these textual memory units into images. However, MemOCR still relies on long-context understanding capabilities to maintain these memory units and neglects the structural information present in raw dialogues.
In contrast to MemOCR, \method{} directly renders raw dialogues into memory snapshots, eliminating the need for additional textual memory construction and preserving both the full information content and structural semantics of the original dialogues.

%% file: sections/formulation.tex
\section{Methodology}

In this section, we first introduce a unified formulation for memory-augmented generation ($\S$~\ref{sec:method:preliminary}), which provides a common framework for understanding how memory assists LLMs in answering questions. We then revisit prior approaches through the lens of text memory and visual memory ($\S$~\ref{sec:method:text_mem}), highlighting their methods in the memory management. Building on this formulation, we finally present \method{} ($\S$~\ref{sec:method:memshot}), which directly renders dialogues into memory units and leverages the visual understanding capabilities of MLLMs for memory-augmented generation modeling.

\input{figures/main}
\subsection{Preliminaries of Memory-Augmented Generation}\label{sec:method:preliminary}
To facilitate long-horizon reasoning capability, LLMs are typically required to answer a given query $q$ by leveraging historical interaction information $\mathcal{C}$, thus handling dynamic and complex environments that involve stateless user interactions:
\begin{equation}\label{eq:rag}
a\sim \text{LLM}(\cdot\mid q, \mathcal{C}),
\end{equation}
where the interaction history $\mathcal{C}$ can be represented as $T$ dialogue turns, potentially involving long-term interactions:
\begin{equation}
\mathcal{C} = \{c_1, c_2, \dots,c_T\}.
\end{equation}
As the dialogue history ($\mathcal{C}$) grows, directly conditioning on the full context becomes increasingly challenging. Due to the lost-in-the-middle problem~\cite{liu2024lost}, long contexts make it difficult for the model to accurately identify relevant information and effectively utilize key knowledge distributed across long-range interactions, which may ultimately degrade answer quality.

To address these issues, existing approaches introduce memory mechanisms that aim to maintain a persistent state, thereby facilitating long-term knowledge utilization by leveraging extended interactions from the dialogue history ($\mathcal{C}$):
\begin{equation}
\mathcal{M} = \Psi(\mathcal{C}),
\end{equation}
where $\Psi(\cdot)$ denotes a memory construction function that extracts salient information from the original dialogue, and $\mathcal{M}$ is the constructed memory set comprising multiple memory units, i.e., $\mathcal{M} = \{M_1, \dots, M_n\}$, which encapsulate additional information derived from the dialogue history to support LLM reasoning.
Different from Eq.~\ref{eq:rag}, the model generates the final answer conditioned on the query $q$ and the constructed memory $\mathcal{M}$:
\begin{equation}\label{eq:memory_gen}
a\sim \text{LLM}(\cdot\mid q, \mathcal{M}).
\end{equation}
To improve the answer generation process in Eq.~\ref{eq:memory_gen}, we can obtain a query-relevant subset $\Tilde{\mathcal{M}}$ to filter out irrelevant memory units:
\begin{equation}
\Tilde{\mathcal{M}}= \text{Retrieval}_{\text{top}-k}(\mathcal{M},q),
\end{equation}
where $\text{Retrieval}_{\text{top}-k}(\cdot)$ denotes a retrieval function that returns the top-$k$ memory units most relevant to the query $q$. 
We then introduce the methodology for constructing the memory using manageable text chunks in $\S$~\ref{sec:method:text_mem}.

\subsection{Constructing Fine-grained Memory via Iterative Text Chunk Updating}\label{sec:method:text_mem}
Existing methods~\cite{hu2026evermemos, li2025memos, kang2025memory} typically define the memory construction function $\Psi(\cdot)$ as an iterative process that unfolds along the temporal order of the dialogue, using it to build manageable text chunks as the memory representation $\mathcal{M}$.
For the constructed memory, existing approaches usually maintain a memory state that iteratively extracts or updates relevant information across all $T$ turns of the dialogue $\mathcal{C}=\{c_1,\dots,c_T\}$, thereby constructing the memory $\mathcal{M}_T$ to support the LLM reasoning process. 

Specifically, let $\mathcal{M}_t$ denote the memory representation after processing the first $t$ dialogue turns. The construction process can then be formulated as:
\begin{equation}
\mathcal{M}_t = \text{Update}(\mathcal{M}_{t-1}, c_t),
\end{equation}
where $\text{Update}(\cdot)$ denotes the memory update function that integrates the current dialogue turn $c_t$ into the existing memory $\mathcal{M}_{t-1}$. $\mathcal{M}_1$ represents the initial state of the memory unit, constructed from the first dialogue turn $c_1$:
\begin{equation}
\mathcal{M}_1 = \text{Initialize}(c_1).
\end{equation}
However, such fine-grained memory construction typically requires iterative updates and may produce fragile memory units that overlook inherent structural information, such as turn boundaries, speaker transitions, and local contextual continuity. Consequently, it becomes difficult to fully exploit knowledge distributed across different dialogue turns, leading existing approaches to repeatedly gather relevant information by re-examining the interaction history, which introduces redundancy and undermines the effectiveness of the constructed memory~\cite{li2025memos, kang2025memory, hu2026evermemos}. To address these limitations, we introduce the \method{} method in $\S$~\ref{sec:method:memshot}, which organizes memory based on dialogue shots as memory units rather than fragmented and independent text chunks.

\subsection{Efficient Memory Construction through Dialogue Chunk Shooting}\label{sec:method:memshot}
To more directly preserve the structural organization of raw dialogue, we introduce \method{}, a dialogue shooting mechanism that constructs structured visual memory units from local contiguous spans of the original dialogue. This design treats the raw dialogue $\mathcal{C}$ as the primary source and renders each local span through hierarchical dialogue templates.

Specifically, we denote the $r$-th dialogue chunk in the original dialogue as:
\begin{equation}
% C^{(r)} = \{c_{r \times t:(r+1) \times t}\},
C^{(r)} = \mathcal{C}_{(r-1)\times t : r\times t},
\end{equation}
where $C^{(r)}$ denotes the $r$-th dialogue chunk and $t$ denotes the turns it contains. Under this formulation, each $C^{(r)}$ remains directly grounded in a temporally localized segment of the original interaction.

\textbf{Global Information Extraction.}
For each local dialogue span $\mathcal{C}^{(r)}$, we organize its content into a hierarchical structure. The upper level captures global cues of the span, while the lower level consists of the utterances within the span, preserving speaker identities, turn order, and local dependencies between adjacent utterances. Thus, we can define a structured dialogue template as:
\begin{equation}
    \tau = \{ \tau_{\text{header}}, \tau_{\text{chat}}\},
\end{equation}
where $\tau_{\text{header}}$ indicates the meta information of the dialogue chunk, and $\tau_{\text{chat}}$ structures the dialogue content into a conversational region.
Concretely, the header region may include session-level metadata such as a session identifier and timestamp (e.g., ``Session 03, May 25, 2023''), providing temporal context. The chat region represents the dialogue as a sequence of utterances, where each message is associated with a speaker (e.g., ``Melanie:'' or ``Caroline:''), and rendered in a speaker-aware layout (e.g., left-right alignment and distinct visual styles). This design preserves explicit turn boundaries and local adjacency, enabling the model to access both semantic content and interaction structure within the span.

\textbf{Structured Memory Shot Rendering.}
After the hierarchical organization, the extracted information is mapped into a unified visual layout. Specifically, $\tau_{\text{header}}$ places session-level information at the top of the shot, while $\tau_{\text{chat}}$ renders the dialogue as a speaker-aware chat stream.
Utterances from different speakers are arranged with consistent relative positions and visual styles, making speaker identity explicit while preserving clear turn boundaries and local adjacency.
Formally, the $r$-th memory shot is defined as:
\begin{equation}
s_r = \Phi \left(\tau_{\text{header}}(\mathcal{C}^{(r)}), \tau_{\text{chat}}(\mathcal{C}^{(r)})\right),
\end{equation}
where $\Phi(\cdot)$ denotes the template rendering function. Rendering all local spans yields a set of visual memory units:
\begin{equation}
\mathcal{M} = \{s_1, s_2, \dots, s_R \},
\end{equation}
where $R = \left\lceil \frac{T}{t} \right\rceil$ denotes the total number of memory shots.
Through hierarchical extraction and structured rendering, each memory shot explicitly preserves the structural properties of its corresponding dialogue span.
As a result, \method{} provides a structured external memory representation that remains closely aligned with the organization of the original dialogue, thereby facilitating more reliable retrieval and reasoning over long-term interaction history.

%% file: figures/main.tex
\begin{figure*}[t]
\centering
\includegraphics[width=\textwidth]{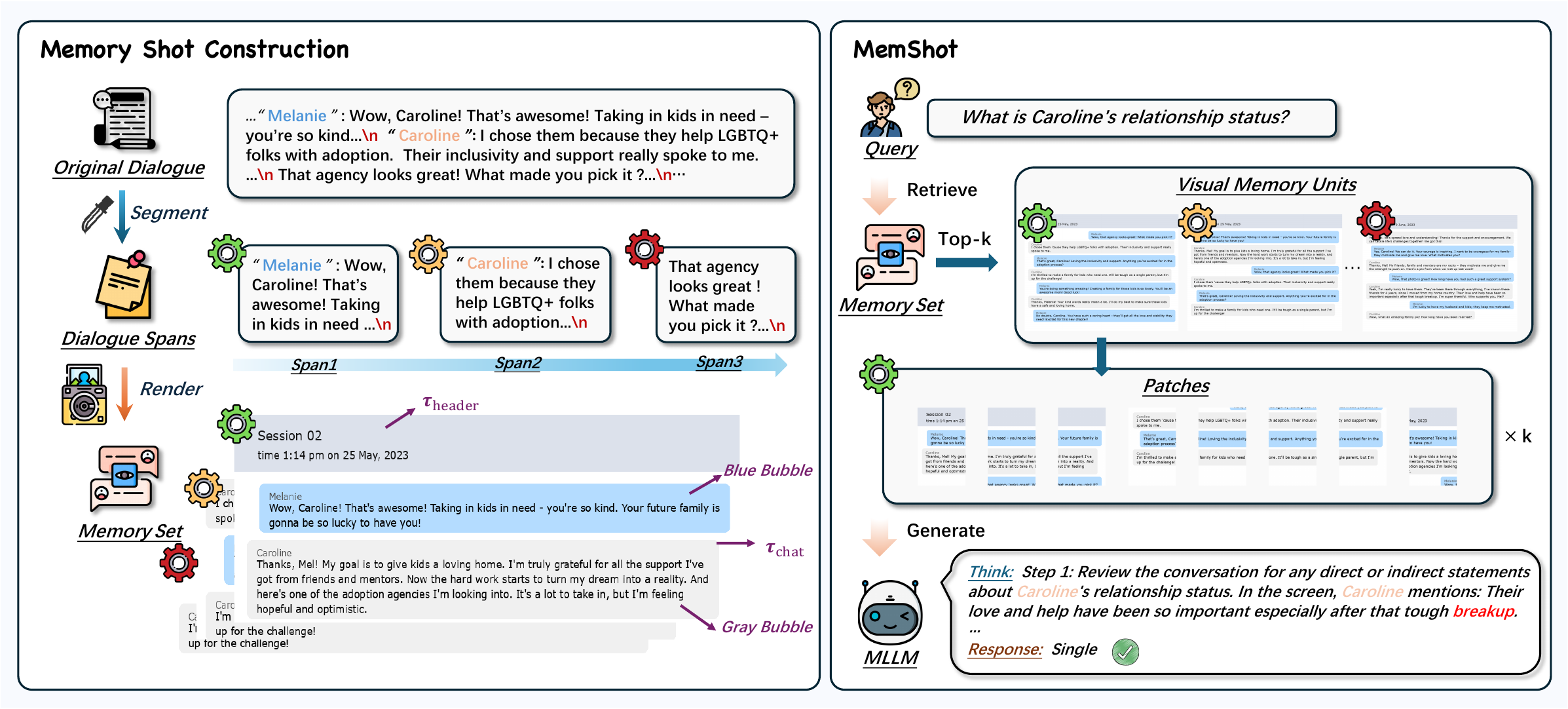}
\caption{Illustration of Our Proposed \method{}.}
\label{fig:main}
\end{figure*}

%% file: sections/experimental_methodology.tex
\section{Experimental Methodology}
This section presents the baselines, datasets, evaluation metrics, and implementation details used in our experiments.

\input{tables/overall_locomo}
\input{tables/overall_longmemeval}

\textbf{Datasets.} Following prior work~\cite{fang2025lightmem,hu2026evermemos}, we conduct experiments on two long-term conversational memory benchmarks, LoCoMo and LongMemEval.
LoCoMo~\cite{maharana2024evaluating} focuses on memory modeling in ultra-long, multi-turn conversations, featuring long-span interaction histories across sessions. LongMemEval~\cite{wu2024longmemeval} emphasizes information extraction, multi-session reasoning, temporal reasoning, and dynamic memory updating in long-term interactions.

\textbf{Baselines.}
To evaluate the effectiveness of visual memory, we compare the Text RAG model and several typical memory models. 

Specifically, the Text RAG model~\cite{ram2023context} treats each dialogue session as a retrieval unit, retrieves relevant text snippets from the conversation history, and directly feeds them into the model for response generation.
We further compare our approach with several representative text-based memory systems. 
LightMem~\cite{fang2025lightmem} conducts a lightweight memory system by organizing historical information through a hierarchical memory construction mechanism. We also include comparisons with MemOS~\cite{li2025memos}, which formulates memory as a first-class system resource and unifies parametric, activation, and textual memories within a schedulable framework, as well as MemoryOS~\cite{kang2025memory}, a hierarchical memory operating system designed to maintain and retrieve memories for persistent and personalized agent behaviors. EverMemOS~\cite{hu2026evermemos} further introduces memory lifecycle modeling to enable more stable long-term memory management.
In addition, we include MemOCR~\cite{shi2026memocr} as a strong visual-memory-related baseline, which first summarizes textual memories and then renders them into visual memory snapshots. More details on baseline implementations are provided in Appendix~\ref{app:baseline_detail}.

\textbf{Evaluation Metrics.}
Following~\citet{fang2025lightmem} and~\citet{xu2025mem}, we adopt both Accuracy and F1 score as evaluation metrics. 
During evaluation, we employ GLM-5~\cite{zeng2026glm} as the judge model, providing it with both the ground-truth answer and the model prediction via the evaluation prompt described in Appendix~\ref{app:prompt}. The judge model is then asked to produce a binary decision indicating whether the prediction is correct, based on which the final Accuracy and F1 scores are computed.

\textbf{Implementation Details.}
In our experiments, \method{} is implemented based on Qwen3-VL-Instruct~\cite{bai2025qwen3} across different scales (2B, 8B, and 32B). For retrieval, we consistently use Qwen3-VL-Embedding-8B~\cite{li2026qwen3} as the retriever, and top-10 ranked memory units are used for generation. 
During visual memory construction, each memory shot is formed by sequentially grouping dialogue turn-pairs until the rendered image reaches the maximum height of 768 pixels. 
To preserve continuity between adjacent shots, when constructing a new shot, we additionally prepend the last two turn-pairs from the previous shot to the current one. 
Additional implementation details are provided in Appendix~\ref{app:impl_detail}.

%% file: tables/overall_locomo.tex
\begin{table*}[t]
\centering
\caption{Overall Performance on LoCoMo Dataset across Different Qwen3 Model Scales. The best results are marked in \textbf{bold}, while the \underline{second-best} results are underlined.}
\label{tab:locomo}
\begin{tabular}{lcrcccccccccc}
\toprule
\multirow{2}{*}{\textbf{Method}} & \multirow{2}{*}{\textbf{Model Scale}} & \multirow{2}{*}{\textbf{Latency (s)}}
& \multicolumn{2}{c}{\textbf{Multi-Hop}}
& \multicolumn{2}{c}{\textbf{Temporal}}
& \multicolumn{2}{c}{\textbf{Open-Domain}}
& \multicolumn{2}{c}{\textbf{Single-Hop}}
& \multicolumn{2}{c}{\textbf{Overall}} \\
\cmidrule(lr){4-5}\cmidrule(lr){6-7}\cmidrule(lr){8-9}\cmidrule(lr){10-11}\cmidrule(lr){12-13}
& & & \textbf{Acc} & \textbf{F1}
& \textbf{Acc} & \textbf{F1}
& \textbf{Acc} & \textbf{F1}
& \textbf{Acc} & \textbf{F1}
& \textbf{Acc} & \textbf{F1} \\
\midrule

\rowcolor{gray!20}\multicolumn{13}{l}{\textbf{Qwen3 (LLM)}} \\
\midrule
Text RAG    & 32B & 2.10    & 59.22 & 27.41 & \textbf{67.91} & \underline{35.81} & 37.50 & 16.80 & 78.00 & 36.81 & \underline{69.94} & 33.63 \\
LightMem~\cite{fang2025lightmem}    & 32B & 814.34  & \underline{60.28} & \underline{37.75} & 55.14 & 31.77 & \underline{51.04} & 21.76 & \underline{80.02} & 49.64 & 69.42 & 42.00 \\
MemOS~\cite{li2025memos}       & 32B & 628.76  & \textbf{69.86} & \textbf{41.31} & \underline{65.11} & 32.99 & \textbf{54.17} & \textbf{29.94} & \textbf{85.26} & \textbf{57.66} & \textbf{76.30} & \textbf{47.79} \\
MemoryOS~\cite{kang2025memory}    & 32B & 2594.09 & 53.55 & 35.65 & 29.91 & \textbf{39.53} & \underline{51.04} & \underline{25.72} & 74.44 & \underline{54.18} & 59.87 & \underline{45.96} \\
EverMemOS~\cite{hu2026evermemos}   & 32B & 1535.98 & 58.87 & 12.62 & 62.31 & 20.20 & \underline{51.04} & 9.30 & 73.01 & 16.94 & 66.82 & 16.35 \\

\midrule
\rowcolor{gray!20}\multicolumn{13}{l}{\textbf{Qwen3-VL-Instruct (MLLM)}} \\
\midrule

Text RAG    & 2B  & 2.10    & 47.16 & 31.31 & 28.66 & 39.18 & 36.46 & 14.56 & 69.44 & 51.08 & 54.81 & 42.70 \\
LightMem~\cite{fang2025lightmem}    & 2B  & 676.03  & \underline{54.96} & \underline{35.15} & \underline{30.53} & \underline{40.48} & \underline{39.58} & \underline{18.23} & 74.91 & 48.96 & 59.81 & 42.75 \\
MemOS~\cite{li2025memos}       & 2B  & 670.60  & \textbf{57.80} & \textbf{37.28} & 28.97 & 33.80 & 38.54 & 17.94 & \underline{76.22} & \underline{54.34} & \underline{60.65} & \underline{44.67} \\
MemoryOS~\cite{kang2025memory}    & 2B  & 571.85  & 39.01 & 15.69 & 12.77 & 34.51 & 32.29 & 7.80 & 58.50 & 23.92 & 43.77 & 23.62 \\
EverMemOS~\cite{hu2026evermemos}    & 2B  & 1670.10 & 37.23 & 8.68 & 25.23 & 16.69 & 29.17 & 4.47 & 60.05 & 11.50 & 46.69 & 11.63 \\
\hdashline
MemOCR~\cite{shi2026memocr}      & 2B  & 3.43    & 28.72 & 26.31 & 15.26 & 25.05 & 37.50 & \textbf{21.40} & 40.43 & 37.54 & 32.86 & 31.87 \\
\method{}   & 2B  & 9.56    & 52.13 & 34.03 & \textbf{37.07} & \textbf{47.21} & \textbf{40.62} & 18.06 & \textbf{81.81} & \textbf{57.57} & \textbf{64.48} & \textbf{48.64} \\

\midrule

Text RAG    & 8B  & 2.10    & 61.70 & \underline{40.24} & 58.57 & \underline{55.68} & 41.67 & 23.87 & 81.09 & 60.09 & 70.39 & \underline{53.28} \\
LightMem~\cite{fang2025lightmem}    & 8B  & 439.06  & 58.16 & 37.85 & 47.98 & 45.84 & 46.88 & 22.08 & 81.33 & 54.32 & 67.99 & 47.53 \\
MemOS~\cite{li2025memos}       & 8B  & 673.58  & \textbf{67.38} & \textbf{44.68} & 45.17 & 39.20 & \underline{50.00} & \underline{25.74} & \underline{83.71} & \underline{61.33} & 70.58 & 51.45 \\
MemoryOS~\cite{kang2025memory}    & 8B  & 1203.24 & 53.90 & 36.17 & 26.48 & 39.66 & 42.71 & 20.89 & 72.77 & 53.41 & 57.79 & 45.36 \\
EverMemOS~\cite{hu2026evermemos}    & 8B  & 929.83  & \underline{66.31} & 15.15 & \underline{68.54} & 22.29 & \underline{50.00} & 8.14 & \textbf{85.37} & 18.98 & \textbf{76.17} & 18.29 \\
\hdashline
MemOCR~\cite{shi2026memocr}        & 8B  & 5.31    & 54.96 & 31.41 & 39.56 & 47.44 & \textbf{57.29} & \textbf{29.85} & 73.84 & 43.35 & 62.21 & 41.17 \\
\method{}   & 8B  & 9.56    & 60.99 & 39.99 & \textbf{70.72} & \textbf{60.07} & 41.67 & 22.84 & \textbf{85.37} & \textbf{64.44} & \underline{75.13} & \textbf{56.46} \\

\midrule

Text RAG    & 32B & 2.10    & 67.73 & \textbf{42.26} & 73.21 & \underline{60.82} & 56.25 & \textbf{31.13} & 84.54 & \underline{61.04} & 77.34 & \underline{55.69} \\
LightMem~\cite{fang2025lightmem}    & 32B & 700.50  & \textbf{70.57} & 38.68 & 59.19 & 49.79 & 50.00 & 20.52 & 84.19 & 57.20 & 74.35 & 49.98 \\
MemOS~\cite{li2025memos}       & 32B & 669.45  & \underline{69.86} & 40.26 & 72.59 & 42.87 & \underline{58.33} & 28.09 & \underline{86.68} & 59.53 & \underline{78.90} & 50.57 \\
MemoryOS~\cite{kang2025memory}    & 32B & 3471.94 & 64.54 & \underline{41.43} & 37.69 & 49.76 & 48.36 & 26.50 & 76.22 & 56.52 & 64.35 & 50.48 \\
EverMemOS~\cite{hu2026evermemos}    & 32B & 1957.94 & 68.79 & 12.16 & \underline{73.52} & 21.48 & 44.79 & 7.42 & 85.14 & 16.00 & 77.21 & 15.91 \\
\hdashline
MemOCR~\cite{shi2026memocr}        & 32B & 12.03   & 65.60 & 29.58 & 67.91 & 57.85 & \textbf{62.50} & 30.05 & 78.36 & 40.03 & 72.86 & 41.21 \\
\method{}   & 32B & 9.56    & 65.96 & 40.62 & \textbf{78.19} & \textbf{63.14} & 53.12 & \underline{30.85} & \textbf{87.75} & \textbf{65.75} & \textbf{79.61} & \textbf{58.43} \\

\bottomrule
\end{tabular}
\end{table*}

%% file: tables/overall_longmemeval.tex
\begin{table*}[t!]
\centering
\caption{Overall Performance on LongMemEval Dataset across Different Qwen3-VL Model Scales. The best results are marked in \textbf{bold}, while the \underline{second-best} results are underlined. 
% The best results are marked in \textbf{bold}, while the \underline{second-best} results are underlined. 
% Acc denotes the accuracy measured by LLM-as-a-Judge. 
% \method{} (Full Session) means that the visual memory of a complete session is used as the retrieval unit.
}
\label{tab:longmemeval}
\resizebox{\linewidth}{!}{
\begin{tabular}{lccccccccccccccc}
\toprule
\multirow{2}{*}{\textbf{Method}}& \multicolumn{2}{c}{\textbf{SS-User}} & \multicolumn{2}{c}{\textbf{SS-Asst}} & \multicolumn{2}{c}{\textbf{SS-pref}} & \multicolumn{2}{c}{\textbf{Multi-S}} & \multicolumn{2}{c}{\textbf{Know. Upd}} & \multicolumn{2}{c}{\textbf{Temp. Reas}} & \multicolumn{2}{c}{\textbf{Overall}} \\
\cmidrule(lr){2-3}\cmidrule(lr){4-5}\cmidrule(lr){6-7}\cmidrule(lr){8-9}\cmidrule(lr){10-11}\cmidrule(lr){12-13}\cmidrule(lr){14-15}
&\textbf{Acc} & \textbf{F1} & \textbf{Acc} & \textbf{F1} & \textbf{Acc} & \textbf{F1} & \textbf{Acc} & \textbf{F1} & \textbf{Acc} & \textbf{F1} & \textbf{Acc} & \textbf{F1} & \textbf{Acc} & \textbf{F1} \\
\midrule

% \rowcolor{gray!20}\multicolumn{16}{l}{\textbf{Qwen3-VL-2B-Instruct}} \\
% \midrule
% Text RAG & 77.14 & 37.68 & 80.36 & 59.44 & 23.33 & \underline{12.51} & 22.56 & 15.65 & \underline{55.13} & 27.82 & 27.82 & 20.61 & 43.20 & 26.67 \\
% LightMem & 55.71 & 13.93 & 16.07 & 6.89 & \textbf{66.67} & 11.80 & 13.53 & 4.78 & 38.46 & 7.74 & 26.32 & 8.15 & 30.20 & 8.08 \\
% % MemU & 85.71 & 45.04 & 67.86 & 29.90 & 53.33 & 11.36 & 42.86 & 16.71 & 56.41 & 25.80 & 30.83 & 18.79 & 51.20 & 23.80 \\
% MemOS & 80.00 & 25.17 & \textbf{92.86} & 27.87 & \underline{53.33} & \textbf{14.19} & \textbf{29.32} & 6.22 & \textbf{56.41} & 12.72 & \textbf{33.83} & 8.61 & \textbf{50.40} & 13.55 \\
% MemoryOS  & 71.43 & 48.21 & 78.57 & 52.59 & 13.33 & 3.39 & \underline{25.56} & 12.58 & 50.00 & 30.79 & 14.29 & 14.99 & 38.00 & 24.98 \\
% EverMemOS & 54.29 & 12.81 & 32.14  & 14.26 & 23.33  & 11.11 & 21.05  & 7.88 & 25.64 & 7.91 & 18.05 & 10.14 & 27.00 & 10.19 \\
% \hdashline
% MemOCR & 21.43 & 18.61 & 23.21 & 22.99 & 13.33 & 9.79 & 9.02 & 8.09 & 37.18 & 31.44 & 18.80 & 19.26 & 19.60 & 17.95 \\
% \method{} (Full Session)  & \underline{84.29} & \underline{70.10} & 89.29 & \underline{74.13} & 30.00 & 9.17 & 21.05 & \underline{20.90} & 46.15 & \textbf{38.64} & \underline{28.57} & \textbf{30.83} & 44.00 & \textbf{38.73} \\
% \method{} & \textbf{91.43} & \textbf{75.38} & \underline{91.07} & \textbf{76.03} & 30.00 & 6.50 & 23.31 & \textbf{21.07} & 48.72 & \underline{37.34} & 26.32 & \underline{27.65} & \underline{45.60} & \underline{38.24} \\

% \midrule
\rowcolor{gray!20}\multicolumn{16}{l}{\textbf{Qwen3-VL-8B-Instruct}} \\
\midrule
Text RAG & 87.14 & 72.59 & \underline{94.64} & \textbf{79.66} & 30.00 & 6.51 & 48.87 & \underline{37.13} & 67.95 & 50.35 & 46.62 & \underline{34.29} & 60.60 & \underline{46.33} \\
LightMem~\cite{fang2025lightmem} & \underline{92.86} & 12.72 & 25.00 & 5.06 & \textbf{80.00} & 9.20 & \textbf{68.42} & 6.63 & 70.51 & 5.66 & \textbf{67.67} & 8.13 & \textbf{67.80} & 7.71 \\
% MemU & 95.71 & 52.62 & 85.71 & 50.32 & 53.33 & 12.19 & 71.43 & 18.67 & 82.05 & 32.32 & 58.65 & 28.82 & 73.60 & 31.41 \\
MemOS~\cite{li2025memos} & 85.71 & 22.80 & \underline{94.64} & 25.48 & \underline{56.67} & \textbf{11.37} & 47.37 & 6.67 & \textbf{80.77} & 7.95 & \underline{51.13} & 7.52 & 64.80 & 11.74 \\
MemoryOS~\cite{kang2025memory}  & 85.71  & \underline{75.94} & 85.71 & 69.51 &  53.33 & 9.85 & 29.32 & 30.72 & 56.41  & \underline{54.08} & 30.08 & 32.91 & 49.40 & 44.37 \\
EverMemOS~\cite{hu2026evermemos} & \underline{92.86} & 30.39 & 67.86 & 27.83 & \underline{56.67} & \underline{10.81} & \underline{60.90} & 11.44 & 74.36 & 17.47 & 49.62 & 15.60 & 65.00 & 17.94 \\
\hdashline
MemOCR~\cite{shi2026memocr} & 88.57 & 67.77 & 89.29 & 63.05 & 33.33 & 8.64 & 42.11 & 31.22 & \underline{79.49} & 49.25 & 29.32 & 27.87 & 55.80 & 40.60 \\
% \method{} (Full Session)  & 90.00 & \underline{77.32} & \underline{94.64} & 77.74 & 36.67 & 6.72 & 48.12 & 34.49 & 61.54 & 49.19 & 47.37 & 33.89 & 60.40 & 45.80 \\
\method{} & \textbf{95.71} & \textbf{82.48} & \textbf{98.21} & \underline{78.05} & 33.33 & 6.94 & 57.89 & \textbf{42.73} & 71.79 & \textbf{61.30} & 48.87 & \textbf{34.88} & \underline{66.00} & \textbf{50.91} \\

\midrule
\rowcolor{gray!20}\multicolumn{16}{l}{\textbf{Qwen3-VL-32B-Instruct}} \\
\midrule
Text RAG & 94.29 & 76.08 & \textbf{98.21} & \textbf{83.62} & 50.00 & 10.41 & 64.66 & \underline{48.60} & 73.08 & 51.07 & \underline{62.41} & 38.93 & 72.40 & 51.89 \\
LightMem~\cite{fang2025lightmem} & \textbf{97.14} & 13.71 & 26.79 & 6.00 & \textbf{93.33} & 11.05 & \textbf{76.69} & 6.49 & 83.33 & 5.79 & \textbf{79.70} & 8.65 & \textbf{76.80} & 8.18 \\
% MemU & 97.14 & 24.31 & 91.07 & 34.06 & 63.33 & 15.49 & 78.20 & 14.06 & 88.46 & 17.67 & 63.91 & 20.20 & 79.20 & 20.02 \\
MemOS~\cite{li2025memos} & 85.71 & 10.24 & \underline{94.64} & 16.69 & 73.33 & \underline{12.86} & 48.12 & 4.76 & \underline{84.62} & 6.73 & 58.65 & 6.27 & 68.60 & 8.06 \\
MemoryOS~\cite{kang2025memory}   & 91.43 & \underline{76.27} & 92.86 & 77.75 & 63.33 & 9.75 & 54.14 & 45.54 & 70.51 & \underline{60.37} & 45.11 & \textbf{42.33} & 64.40 & \underline{52.76} \\
EverMemOS~\cite{hu2026evermemos} & 94.29 & 26.06 & 89.29 & 25.95 & \underline{90.00} & \textbf{13.55} & 64.66 & 10.64 & 83.33 & 14.61 & 57.89 & 15.98 & 74.20 & 16.73 \\
\hdashline
MemOCR~\cite{shi2026memocr} & 85.71 & 70.53 & \textbf{98.21} & 69.20 & 50.00 & 11.98 & 52.63 & 40.89 & 76.92 & 53.69 & 36.84 & 27.90 & 61.80 & 45.32 \\
% \method{} (Full Session) & \underline{95.71} & \textbf{82.57} & 92.86 & 78.50 & 53.33 & 10.91 & \underline{70.68} & \textbf{57.27} & 83.33 & 58.09 & \underline{63.91} & \textbf{45.34} & \underline{75.80} & \textbf{57.36} \\
\method{} & \underline{95.71} & \textbf{81.76} & \textbf{98.21} & \underline{82.51} & 50.00 & 8.47 & \underline{66.17} & \textbf{51.47} & \textbf{89.74} & \textbf{61.49} & 59.40 & \underline{41.53} & \underline{74.80} & \textbf{55.53} \\

\bottomrule
\end{tabular}}
\end{table*}

%% file: sections/experiments.tex
\section{Evaluation Results}

In this section, we first present the overall performance of \method{} on the LoCoMo and LongMemEval datasets. 
We then conduct a series of analyses to examine the effectiveness of the memory units constructed by \method{} in enhancing MLLMs.

\subsection{Overall Performance} \label{sec:overall}
In this subsection, we compare \method{} against a diverse set of text-based memory baselines on two long-term conversational question answering benchmarks, LoCoMo and LongMemEval, to assess the effectiveness of the visual memory constructed by \method{}.

As shown in Table~\ref{tab:locomo}, we first present the performance of \method{} on the LoCoMo dataset. Overall, \method{} achieves superior performance compared to strong text-based memory systems, while requiring substantially lower memory construction latency, demonstrating both effectiveness and efficiency. 
Unlike text-based memory methods, \method{} disentangles heavy-weight memory-augmented generation models from the need to assimilate associated evidence from user interactions by constructing visual memories from raw dialogs. 
We further compare models across different scales of Qwen3-VL-Instruct. Among these, text-based memory methods typically achieve the lowest memory construction latency with 8B-scale backbones. This may be because 2B models require more inference steps during memory construction due to limited capability, whereas 32B models incur higher latency due to their larger parameter size. 
In contrast, \method{} maintains nearly constant memory construction latency across model scales, and the same constructed memory units preserve effectiveness when applied to different MLLMs, demonstrating strong generalization capability.

\input{tables/ablation}
As shown in Table~\ref{tab:longmemeval}, we further report the performance of \method{} on the LongMemEval dataset. 
The results show that \method{} again achieves competitive performance compared to strong text-based memory systems across both 8B and 32B backbones\footnote{We include the Qwen3-VL-2B-Instruct results on LongMemEval in Appendix~\ref{app:perf_lme_2b}.}, further validating its effectiveness. 
The advantage is particularly evident in categories such as SS-User, SS-Asst, Multi-S, and Knowledge Update scenarios, where modeling user-specific information and cross-session dependencies is critical. 
These results suggest that the structure-preserving visual memory is effective not only within a single scenario but also across diverse long-term conversational settings. 
Notably, compared with the visual memory-based method MemOCR, \method{} achieves improvements of over 10\%, indicating that text-based memory construction may hinder the visual understanding capability of MLLMs. 
In contrast, \method{} better leverages the inherent visual perception and reasoning abilities of MLLMs by directly structuring and rendering raw dialogue chunks into visual memory representations.

\input{sections/mechanism1}
\input{sections/mechanism2}

\input{sections/mechanism3}

%% file: tables/ablation.tex
\begin{table}[t]
\centering
\caption{Ablation Study. All experiments are conducted on the LoCoMo dataset.}
\label{tab:ablation}
\begin{tabular}{lccccc}
\toprule
\multirow{2}{*}{\textbf{Model}} 
& \multicolumn{2}{c}{\textbf{Top-5}} 
& \multicolumn{2}{c}{\textbf{Top-10}} \\
\cmidrule(lr){2-3}\cmidrule(lr){4-5}
& \textbf{Acc} & \textbf{F1} & \textbf{Acc} & \textbf{F1} \\
\midrule

\rowcolor{gray!20}\multicolumn{5}{l}{\textbf{Qwen3-VL-2B-Instruct}} \\
\midrule
\method{} (Full Session) & 58.83 & 44.12 & 59.55 & 44.64 \\
w/o Rendering &54.61& 42.10 & 54.81  & 42.70\\
\hdashline
\method{} & \textbf{64.22} & \textbf{48.67} & \textbf{64.48} & \textbf{48.64} \\
w/o Header & \underline{63.25} & \underline{48.16} & \underline{63.83} & \underline{47.37} \\
w/o Rendering &59.42 & 44.76 & 62.47  & 44.21 \\
%w/o information extractor & 62.14 & 48.16 & 62.34 & 48.05 \\
% w/o bubble color & \underline{63.70} & 48.33 & 63.25 & 47.79 \\
% w/o dialogue bubble & 63.38 & 48.11 & 63.57 & 48.62 \\

\midrule
\rowcolor{gray!20}\multicolumn{5}{l}{\textbf{Qwen3-VL-8B-Instruct}} \\
\midrule
\method{} (Full Session) & 70.26 & 52.93 & \underline{74.87} & 54.11 \\
w/o Rendering &67.86& 51.89 & 70.39  &  53.28\\
\hdashline
\method{} & \textbf{72.73} & \underline{55.11} & \textbf{75.13} & \textbf{56.46} \\
w/o Header & \underline{70.71} & \textbf{55.78} & 74.22 & 55.54 \\
w/o Rendering & 67.47& 52.52&71.62 &\underline{56.01}\\
%w/o information extractor & -- & 54.77 & -- & 55.67 \\
% w/o bubble color & -- & 55.28 & -- & -- \\
% w/o dialogue bubble & -- & 55.18 & -- & -- \\

\bottomrule
\end{tabular}
\end{table}

% \begin{table}[t]
% \centering
% \caption{Ablation Study. 
% Top-k Acc. denotes the answer accuracy conditioned on the top-k retrieved evidence, while Similarity measures the average question-image similarity over the annotated evidence.
% All experiments are conducted using the Qwen3-VL-2B-Instruct model on the LoCoMo dataset.}
% \label{tab:ablation}
% \resizebox{\linewidth}{!}{
% \begin{tabular}{lccc}
% \toprule
% \textbf{Model} & \textbf{Similarity} & \textbf{Top-5 Acc.} & \textbf{Top-10 Acc.} \\
% \midrule
% \method{} & \textbf{47.36} & \textbf{64.22} & \textbf{64.81} \\
% \hdashline
% w/o meta information & 45.61 & 63.25  & \underline{63.83} \\
% w/o information extractor & 44.99 & 62.14 & 62.34 \\
% w/o bubble color & \underline{46.78} & \underline{63.70} & 63.25\\
% w/o dialogue bubble & 45.78 & 63.38 & 63.57 \\
% \bottomrule
% \end{tabular}}
% \end{table}

%% file: sections/mechanism1.tex
\subsection{Ablation Studies}
\label{sec:ablation}
To further differentiate strategies used in \method{}, we conduct ablation studies on different components of the visual memory construction.

As shown in Table~\ref{tab:ablation}, we first evaluate \method{} (Full Session) to examine the effect of different chunking strategies. Unlike \method{}, \method{} (Full Session) segments the dialogue based on session information, which contains richer contextual information than the fixed chunking strategy adopted by \method{}. We then remove the header information and use only the plain text to represent memory units, resulting in two ablation variants: \method{} (w/o Header) and \method{} (w/o Rendering).

\input{figures/recall_and_acc}
As indicated by the evaluation results, \method{} generally achieves better performance than \method{} (Full Session), particularly for smaller-scale MLLMs. A plausible explanation is that although \method{} (Full Session) incorporates more information into each memory unit, it produces larger rendered images, which may hinder comprehension for MLLMs, especially those with limited capacity.
We further analyze the impact of header information. When more retrieved memory units are incorporated (from Top-5 to Top-10), \method{} (w/o Header) exhibits a slight performance decline despite the introduction of additional relevant knowledge, suggesting that these visual memory units increasingly act as noise during generation. In contrast, after incorporating header information, \method{} consistently improves when using Top-10 retrieved memory units compared to Top-5, and outperforms \method{} (w/o Header) by more than 1\%. This demonstrates the effectiveness of header information in enhancing the ability of MLLMs to organize and utilize retrieved memory units.
Finally, compared with \method{}, \method{} (w/o Rendering) suffers the most significant performance drop, indicating that visually structured memory units facilitate more effective knowledge assimilation, even when both textual and visual memory units contain identical information.

%% file: figures/recall_and_acc.tex
\begin{figure}[t]
\centering
\subfigure[Text RAG.]{
\includegraphics[width=0.46\linewidth]{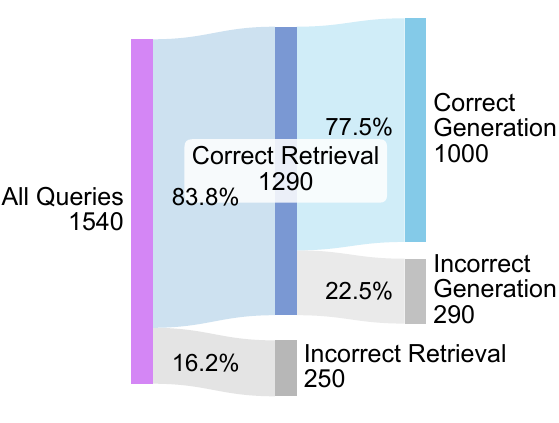}
\label{fig:recall_acc:a}
}
\subfigure[\method{}.]{
\includegraphics[width=0.46\linewidth]{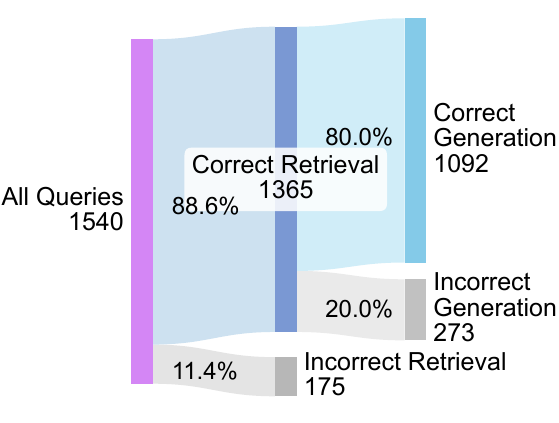}
\label{fig:recall_acc:b}
}
\caption{Performance of Text RAG and \method{} on the LoCoMo dataset. For all methods, the top-10 retrieved memory units are used to augment Qwen3-VL-8B-Instruct.}
\label{fig:recall_acc}
\vspace{-0.1in}
\end{figure}

%% file: sections/mechanism2.tex
\subsection{The Impact of Retrieval with Memory Units Constructed by \method{}}
To investigate the effectiveness of \method{}, we first analyze retrieval performance and robustness based on the visual memory units constructed by \method{}.

\input{figures/recall_analysis}
As shown in Figure~\ref{fig:recall_acc}, we illustrate the interaction between retrieval and generation by decomposing the end-to-end performance into retrieval correctness and generation fidelity. Overall, the evaluation results demonstrate that \method{} mitigates error propagation along the retrieval-augmented generation pipeline, highlighting its effectiveness. Specifically, \method{} significantly improves retrieval accuracy, achieving approximately 5\% gains under the same retrieval budget (top-10), which indicates its superior ability to identify relevant memory units. Furthermore, conditioned on correctly retrieved memory units, \method{} yields more than 2\% improvement over the text-based RAG method, suggesting that the advantages of visual memory units can be effectively translated into improved generation performance via a higher retrieval-to-generation conversion rate.

We further examine the advantages of visual memory units in retrieval by comparing \method{} with a text-based RAG model using two retrieval models, Qwen3-VL-Embedding-8B~\cite{li2026qwen3} and Jina-Embeddings-v4-4B~\cite{gunther2025jina}, as shown in Figure~\ref{fig:recall_analysis}. In Figure~\ref{fig:recall_analysis:a}, we report the recall scores of these retriever families when retrieving memory units constructed from textual and visual representations. The results show that visual memory units are more easily recalled across different retrievers, further validating the effectiveness of \method{}. Notably, the improvements brought by visual memory units are more pronounced when using a relatively weaker embedding model, Jina-Embeddings-v4-4B, indicating that \method{} generalizes well across different retrieval systems. Then we also evaluate the retrieval robustness of \method{} in Figure~\ref{fig:recall_analysis:b}. The results show that \method{} consistently outperforms text-based memory across varying unit sizes, demonstrating that its effectiveness is maintained under different chunking strategies. In contrast, text-based RAG methods are more sensitive to chunk size, which has motivated extensive prior work~\cite{bhat2025rethinking} on optimal chunking strategies. \method{} alleviates this sensitivity by achieving stable retrieval recall after truncating each memory unit to four turns, further highlighting its robustness.

%% file: figures/recall_analysis.tex
\begin{figure}[t]
\centering
\subfigure[Retrieval Performance.]{
\includegraphics[width=0.46\linewidth]{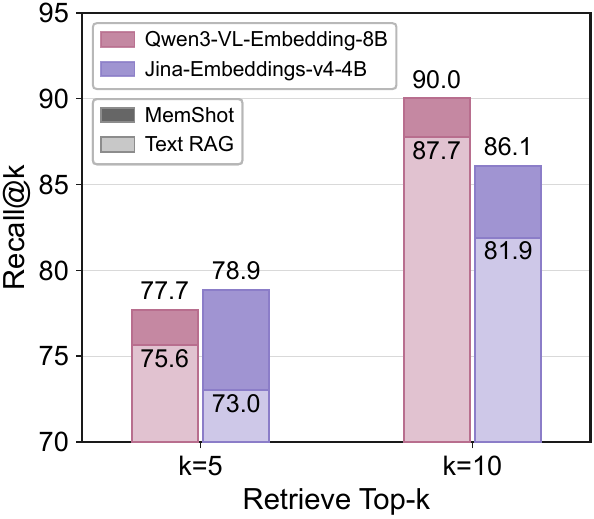}
\label{fig:recall_analysis:a}
}
\subfigure[Retrieval Robustness.]{
\includegraphics[width=0.46\linewidth]{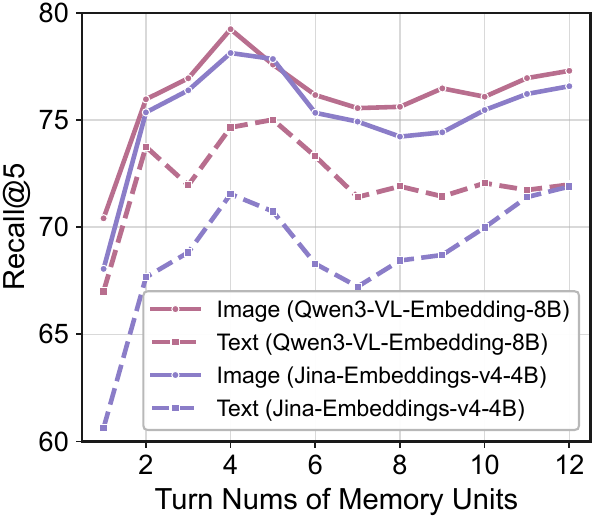}
\label{fig:recall_analysis:b}
}
\caption{Retrieval Performance of Memory Units Constructed by Text RAG and 
\method{} Models on the LoCoMo Dataset.}
\label{fig:recall_analysis}
\end{figure}

%\subfigure[Relationship between text retrieval difficulty and the relative accuracy gain brought by visual memory.]{
%\includegraphics[width=0.46\linewidth]{images/vis_gain.pdf}
%\label{fig:recall_analysis:b}
%}

%\subfigure[Robustness across different memory window sizes using different multimodal retrievers.]{
%\includegraphics[width=0.46\linewidth]{images/recall_text_image_2.pdf}
%\label{fig:recall_analysis:d}
%}

%% file: sections/mechanism3.tex
\subsection{Effectiveness of \method{} in Augmenting MLLM Reasoning}
To analyze how the memory units constructed by \method{} enhance the reasoning capabilities of MLLMs, we evaluate their effectiveness from two complementary perspectives: memory-augmented reasoning and implicit evidence attribution. 

\input{figures/generation}
\textbf{Generation Quality of MLLMs Augmented by \method{}.}
To assess the effectiveness of \method{} in supporting MLLM reasoning, we examine the generation quality of MLLMs conditioned on memory units constructed by Text RAG and \method{}, which represent dialogue snapshots in different formats.

During MLLM response quality evaluation, we prompt MLLMs to generate Chains-of-Thought (CoT)~\cite{wei2022chain} based on the provided memory units, and then conduct a rubric-based LLM-as-a-Judge evaluation~\cite{hashemi2024llm} to assess the generated outputs based on different memory units. 
During evaluation, we provide GLM-5~\cite{zeng2026glm} with the corresponding memory inputs, generated CoTs, and final answers for each method, and assess their memory utilization quality along six dimensions: Structural Information (Structural), Temporal and Dialogue Order Accuracy (Temporal), Conflict Resolution over Memory Evidence (Conflict), Completeness of Consideration (Completeness), Evidence Grounding (Grounding), and Uncertainty Handling and Calibration (Uncertainty).
As shown in Figure~\ref{fig:rubric}, the evaluation results indicate that \method{} consistently outperforms Text RAG across all six dimensions. This suggests that \method{} more effectively enables MLLMs to capture critical reasoning cues through structured visual memory units. Moreover, \method{} demonstrates larger gains in the Uncertainty and Completeness dimensions. This improvement likely stems from the fact that \method{} avoids the brute-force text chunking strategy of Text RAG and instead leverages image rendering to preserve richer structural information, resulting in more self-contained memory units and thereby reducing potential misunderstandings compared to text-based chunks.

\input{figures/case}
\textbf{Evidence Attribution.}
To further investigate whether \method{} alters how the model reads long-term memory from the perspective of implicit evidence attribution, we follow~\citet{fan2025improving} and adopt saliency scores to quantify the contribution of input evidence.

Specifically, for each attention head $h$ in layer $l$, the saliency score between tokens with respect to answer $a$ is defined as:
\begin{equation}
S(a)=\left| \sum_h\left(A^{h,l}\odot\frac{\partial \mathcal{L}(a)}{\partial A^{h,l}} \right)\right|,
\end{equation}
where $A^{h,l}$ is the attention matrix at head $h$ and layer $l$, $\odot$ represents element-wise multiplication, and $\mathcal{L}(a)$ is the cross-entropy loss.

Based on this metric, Figure~\ref{fig:saliency_diff} compares the distributions of saliency scores under text memory and \method{}, illustrating how the two memory interfaces differ in evidence attribution. We observe that, compared to text memory, \method{} yields a more concentrated distribution, whereas text memory exhibits a long-tailed distribution with a larger proportion of tokens receiving negative saliency scores. This pattern suggests that evidence attribution in visual memory is more stable and less noisy, while flattened text memory is more prone to producing dispersed saliency scores, which may mislead MLLMs. This advantage likely stems from the higher information density of visual memory units, which enables them to aggregate more useful knowledge from raw dialogue, thereby making them more effective for supporting answer generation.

\subsection{Case Study}
For qualitative analysis, we randomly select a representative case to examine the evidence attribution behaviors of Text RAG and \method{} in Figure~\ref{fig:case}.

As shown in the evaluation results, the text memory leads MLLMs to produce an incorrect answer, ``married''. Its high-saliency regions are relatively evenly distributed across the flattened transcript, without forming a clearly focused evidence span. This pattern suggests that the model fails to concentrate on a decisive cue and is instead influenced by surface-level lexical signals scattered throughout the text, such as the token ``married'', which appears when Caroline inquires about Malanie's relationship status.
In contrast, \method{} correctly predicts the answer ``Single'', and its saliency map exhibits a much more concentrated attribution pattern. The highlighted regions are primarily localized within the dialogue segment where Caroline mentions a ``tough breakup'', indicating that the MLLM successfully identifies more direct and relevant historical evidence based on these visual memory units. Compared with the diffuse attribution observed in text memory, \method{} demonstrates more localized and selective evidence attribution. This finding further suggests that preserving local dialogue structure through visual representations enables the MLLM to form a clearer evidence focus, ultimately leading to more reliable predictions.

%% file: figures/generation.tex
\begin{figure}[t]
\centering
\subfigure[The Quality of Memory-Augmented Generation Results.]{
\includegraphics[width=0.47\linewidth]{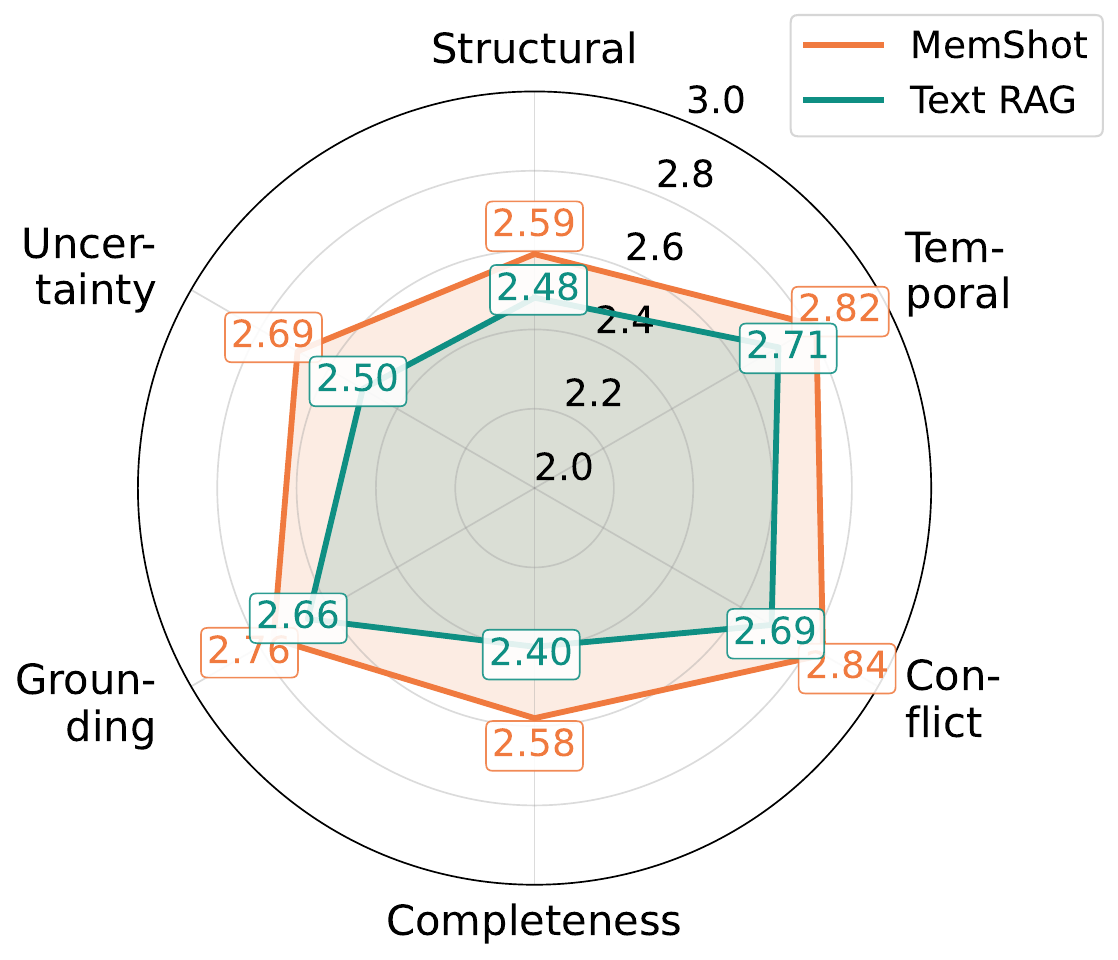}
\label{fig:rubric}
}
\subfigure[Saliency Score Distribution.]{
\includegraphics[width=0.47\linewidth]{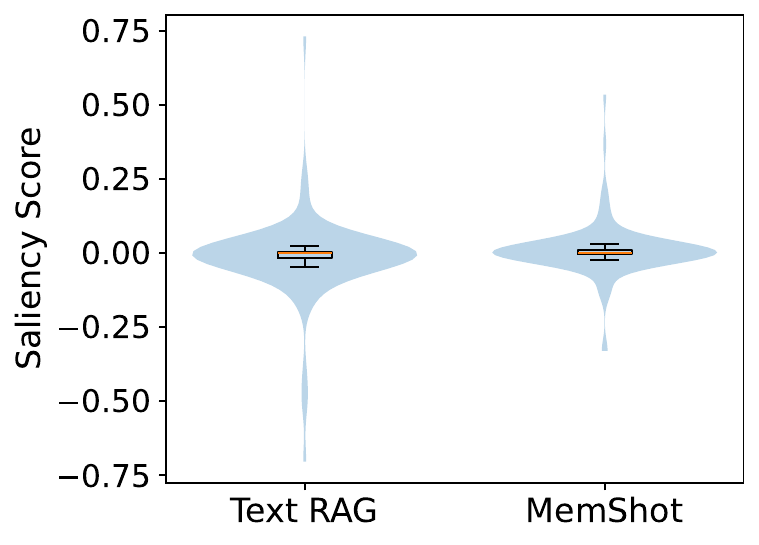}
\label{fig:saliency_diff}
}
\caption{Memory-Augmented Generation of MLLMs Using Memory Units Produced by Text RAG and \method{}.}
\label{fig:memory_reading}
\end{figure}

%% file: figures/case.tex
\begin{figure}[t]
\includegraphics[width=\linewidth]{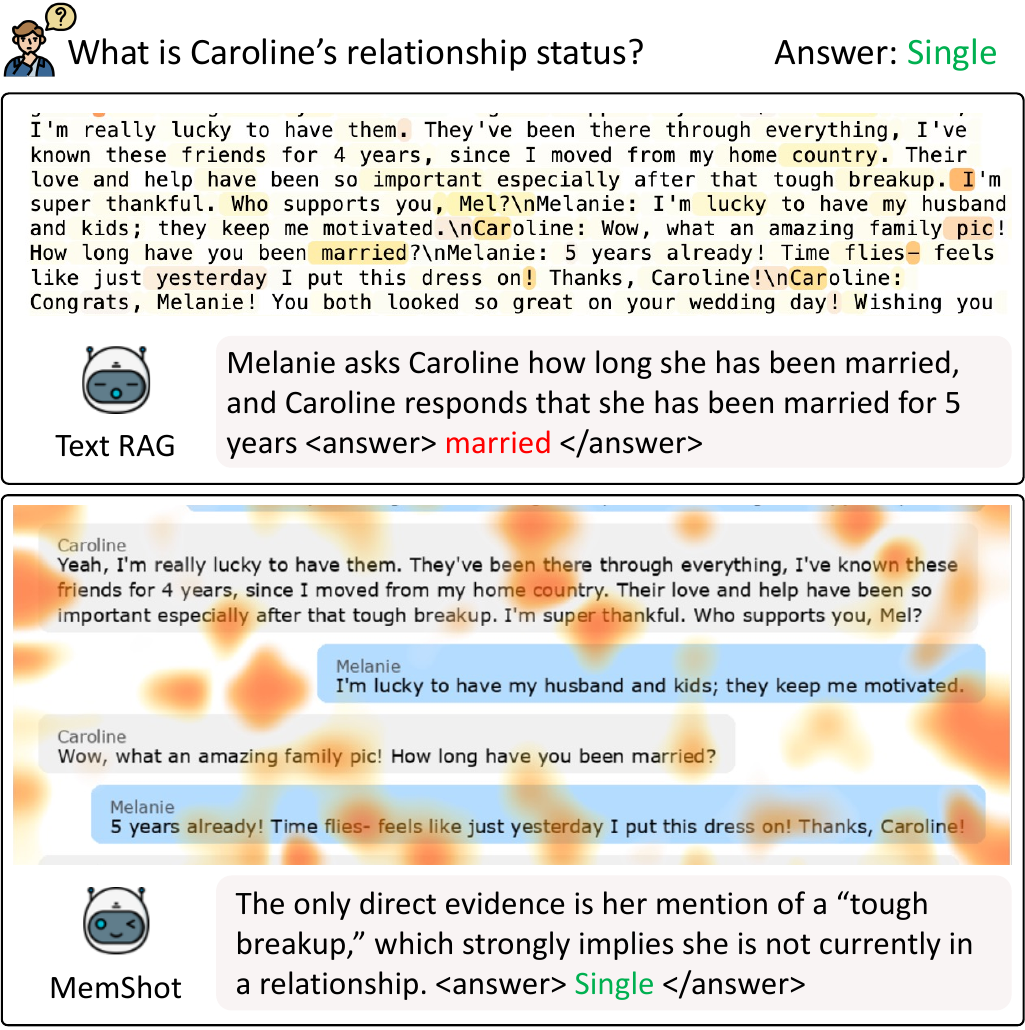}
\caption{Saliency Scores of the Input Evidence for Supporting the Generated Answer. \textcolor{orange}{Orange} highlights evidence with positive saliency, with darker shades indicating greater contribution to the predicted answer.}
\label{fig:case}
\end{figure}

%% file: sections/conclusion.tex
\section{Conclusion}
We propose \method{}, a direct visual memory framework for long-term dialogue that constructs structured memory shots from local contiguous dialogue spans, eliminating the need for sophisticated text-based memory construction pipelines. Experiments on LoCoMo and LongMemEval demonstrate that \method{} achieves stable and competitive performance while substantially simplifying the memory construction process, yielding up to 70$\times$ faster construction. Further analysis shows that \method{} improves the retrieval and utilization of evidence in long-term dialogues, while enabling a more effective memory-augmented generation paradigm through its visual memory construction.

%% file: sections/appendix.tex
\section{Appendix}

\subsection{License}
We strictly comply with the original licenses and usage terms of all datasets used in this work and do not redistribute any third-party raw image content. 
Among the datasets considered in this paper, LoCoMo~\cite{maharana2024evaluating} is released under the Creative Commons Attribution-NonCommercial 4.0 International (CC BY-NC 4.0) license, and LongMemEval~\cite{wu2024longmemeval} is released under the MIT License. 
We use only the officially released versions of these datasets in accordance with their original terms.
In particular, for LoCoMo, the official release does not provide the original images directly, but only image URLs, image descriptions, and search queries when available. 
Accordingly, we do not redistribute any unreleased third-party raw images. 
The visual memory images used in our method are generated renderings derived from the released dialogue text and metadata, rather than copies of third-party raw image content.

\subsection{Implementation Details of Baseline Methods} \label{app:baseline_detail}
We evaluate all baselines under a unified experimental setting. Whenever the original pipeline allows such substitution, we use Qwen3-VL-Instruct as the backbone model for final answer generation, and Qwen3-VL-Embedding-8B~\cite{li2026qwen3} as the shared retriever for both text and image retrieval. For each baseline, we follow the officially released codebase and scripts whenever available, and introduce only minimal compatibility modifications when the original implementation does not directly support a target dataset or evaluation interface.

On LoCoMo, all text-based baselines considered in this work provide official reproduction scripts, which we follow directly in our experiments. On LongMemEval, MemOS~\cite{li2025memos}, LightMem~\cite{fang2025lightmem}, and EverMemOS~\cite{hu2026evermemos} are reproduced using their officially released dataset adaptation scripts. Since MemoryOS~\cite{kang2025memory} does not provide an official adaptation pipeline for LongMemEval, we convert LongMemEval into the LoCoMo-style format following its released conversion script, and then run the official MemoryOS pipeline on the converted data.

MemOCR~\cite{shi2026memocr} is handled separately, as its official implementation does not natively support either LoCoMo~\cite{maharana2024evaluating} or LongMemEval~\cite{wu2024longmemeval}. We therefore adapt both datasets by following their original rendering pipeline as closely as possible. In particular, we preserve its summarize-then-render procedure and modify only the data preprocessing interface required for compatibility with our evaluation setup, without changing its core memory construction logic. For deployment, all generation models and retrievers are served with vLLM\footnote{\url{https://vllm.ai/}} in Docker-based environments.

\subsection{Additional Experimental Details of Visual Memory Rendering} \label{app:impl_detail}
Following the dialogue-shot construction, we render each conversation session into a sequence of structured visual memory units. 
Each unit corresponds to a local contiguous dialogue span and is organized with a unified hierarchical template consisting of a header region and a chat region. 
The header presents session-level meta-information, including the session identifier and timestamp, while the chat region renders utterances in a speaker-aware layout to preserve turn boundaries, speaker positions, and local adjacency between neighboring utterances. 
For all experiments, the image width is fixed at 948 pixels, and the target image height is set to 768 pixels. 
The detailed rendering configuration and an example visual memory unit are shown in Figure~\ref{fig:render_config}.

\input{figures/render_config}

Under this fixed-height setting, we sequentially pack complete turn-pairs into each visual memory unit until the rendered content reaches the height limit. 
When a session is divided into multiple units, each new unit is constructed by first prepending the last two turn-pairs from the previous unit whenever possible, in order to preserve local continuity across adjacent memory shots. 
For LoCoMo~\cite{maharana2024evaluating}, we follow this overlap-by-two strategy by default, and reduce the overlap only when the rendered content cannot fit within the 768-pixel constraint. 
For LongMemEval~\cite{wu2024longmemeval}, since individual turn-pairs are often longer, we keep the same packing strategy but allow the image height to expand when a single complete turn-pair cannot be accommodated within the fixed limit. 
In this way, each visual memory unit remains directly grounded in a temporally localized dialogue span while preserving the structural cues that \method{} relies on for retrieval and downstream reasoning.

\input{tables/overall_longmemeval_2b}

% \textbf{Effect of Maximum Rendering Height.}
\input{tables/render_height}
As shown in Table~\ref{tab:rendering_height}, we further examine how the maximum rendering height affects the quality of the visual memory units on LoCoMo with Qwen3-VL-8B-Instruct~\cite{bai2025qwen3}. Among all evaluated settings, the 768-pixel configuration achieves the best overall performance under both Top-5 and Top-10 retrieval, indicating that the effectiveness of visual memory depends not only on dialogue segmentation but also on the granularity of the rendered memory units. Full-session rendering preserves richer within-session context, but its coarser-grained visual units may reduce retrieval flexibility and make downstream utilization more difficult. In contrast, the 512-pixel setting yields more fragmented memory shots, which may weaken local semantic continuity and structural coherence. Although the 1024-pixel setting remains competitive on some metrics, it still underperforms the 768-pixel configuration overall. We therefore adopt 768 pixels as the default maximum rendering height in MemShot.

\subsection{Additional LongMemEval Results with Qwen3-VL-2B-Instruct}
\label{app:perf_lme_2b}
Table~\ref{tab:longmemeval_2b} reports the additional results on LongMemEval using Qwen3-VL-2B-Instruct as the backbone model. 

Consistent with the main results in Section~\ref{sec:overall}, \method{} remains competitive under the small-model setting and achieves the best overall F1 score among all compared methods. This result suggests that the effectiveness of \method{} does not depend on large model scale alone. Even with a lightweight MLLM, the proposed visual memory design can still provide useful structural cues for retrieval and response generation, leading to strong performance on long-term dialogue understanding. More broadly, these results indicate that the advantage of \method{} comes from the memory interface itself, rather than from relying on stronger large-scale model capacity, and that the proposed approach generalizes well across different backbone sizes.

\subsection{Instruction Prompts} \label{app:prompt}
This subsection summarizes the prompt templates used throughout our evaluation and inference pipeline. 

Figure~\ref{fig:glm5_judge_prompt} shows the prompt used for the standard LLM-as-Judge evaluation with GLM-5~\cite{zeng2026glm}, which is used to determine answer correctness in the main experiments. Figure~\ref{fig:text_rag_prompt} presents the inference prompt for the Text RAG baseline, where the model is asked to answer questions based on retrieved textual memory units. Figure~\ref{fig:memshot_prompt} shows the corresponding inference prompt for \method{}, where the model instead reasons over retrieved visual memory units rendered from dialogue sessions. Together, these prompts illustrate how we keep the answer generation setting aligned across baselines while adapting the input format to their respective memory representations.

Figure~\ref{fig:rubric_cot_analysis_prompt} further presents the prompt used in our rubric-based CoT analysis. Unlike the standard answer-level judge prompt, this prompt is designed to evaluate how well the model utilizes the provided memory during reasoning, rather than only whether the final answer is correct. Specifically, we provide the judge model with three components: the memory input used for generation, the model-generated Chain-of-Thought (CoT)~\cite{wei2022chain}, and the final answer. The judge is then asked to assess the reasoning quality along six rubric dimensions: Structural Information, which measures whether the reasoning preserves speaker identities, turn boundaries, and local dialogue organization; Temporal and Dialogue Order Accuracy, which examines whether the model correctly tracks timestamps and the chronological order of events; Conflict Resolution over Memory Evidence, which evaluates whether the reasoning can reconcile potentially competing or confusing evidence; Completeness of Consideration, which measures whether the model considers all critical clues instead of relying on a partial cue; Evidence Grounding, which checks whether the provided memory explicitly supports the reasoning; and Uncertainty Handling and Calibration, which assesses whether the model expresses appropriate confidence when the evidence is insufficient or ambiguous. This rubric-based analysis allows us to compare Text RAG and \method{} not only at the level of final answer quality, but also in terms of how effectively each memory format supports structured, grounded, and temporally coherent reasoning.

\input{figures/llm_judge_prompt}
\subsection{Additional Case Studies}
Figure~\ref{fig:appendix_case1} shows a representative example of temporal reasoning in long-term dialogue. To answer the question correctly, the model must jointly use the session timestamp and the temporally grounded utterance ``next month.'' Under Text RAG, the flattened text memory weakens this local temporal structure, causing the model to anchor its reasoning to the dialogue date and incorrectly predict ``September 2023.'' By contrast, \method{} preserves both session-level meta-information and turn-level locality in a structured visual memory unit, enabling the model to associate the timestamp with the relevant utterance and correctly infer ``October 2023.'' This example supports our central motivation that preserving dialogue structure, rather than compressing history into fragile text memory, leads to more reliable use of historical evidence in long-term dialogue reasoning.

Figure~\ref{fig:appendix_case2} further presents a representative example of multi-session evidence aggregation. To answer the question correctly, the model must identify and combine two road trip events mentioned across different sessions in May 2023: the Rockies trip in Session 01 and the Jasper trip in Session 02. Under Text RAG, the flattened text memory weakens the separation between local events and nearby dialogue content, causing the model to focus on the explicitly labeled Jasper road trip while overlooking that the Rockies trip is also a valid road trip event, and thus incorrectly predicts only one trip. In contrast, MemShot preserves each dialogue span as a structured visual memory unit with clear session boundaries and localized event descriptions, allowing the model to align both trips with their corresponding timestamps and aggregate them correctly. This case further supports our motivation that preserving dialogue structure and local coherence is more effective than relying on fragile, flattened text memory, especially when the model must compose evidence across multiple sessions rather than match isolated lexical cues.

\input{figures/appendix_case1}

\input{figures/appendix_case2}

\input{figures/text_rag_prompt}

\input{figures/memshot_prompt}

\input{figures/rubric_cot_analysis_prompt}

%% file: figures/render_config.tex
\begin{figure}[t]
\centering
\includegraphics[width=\linewidth]{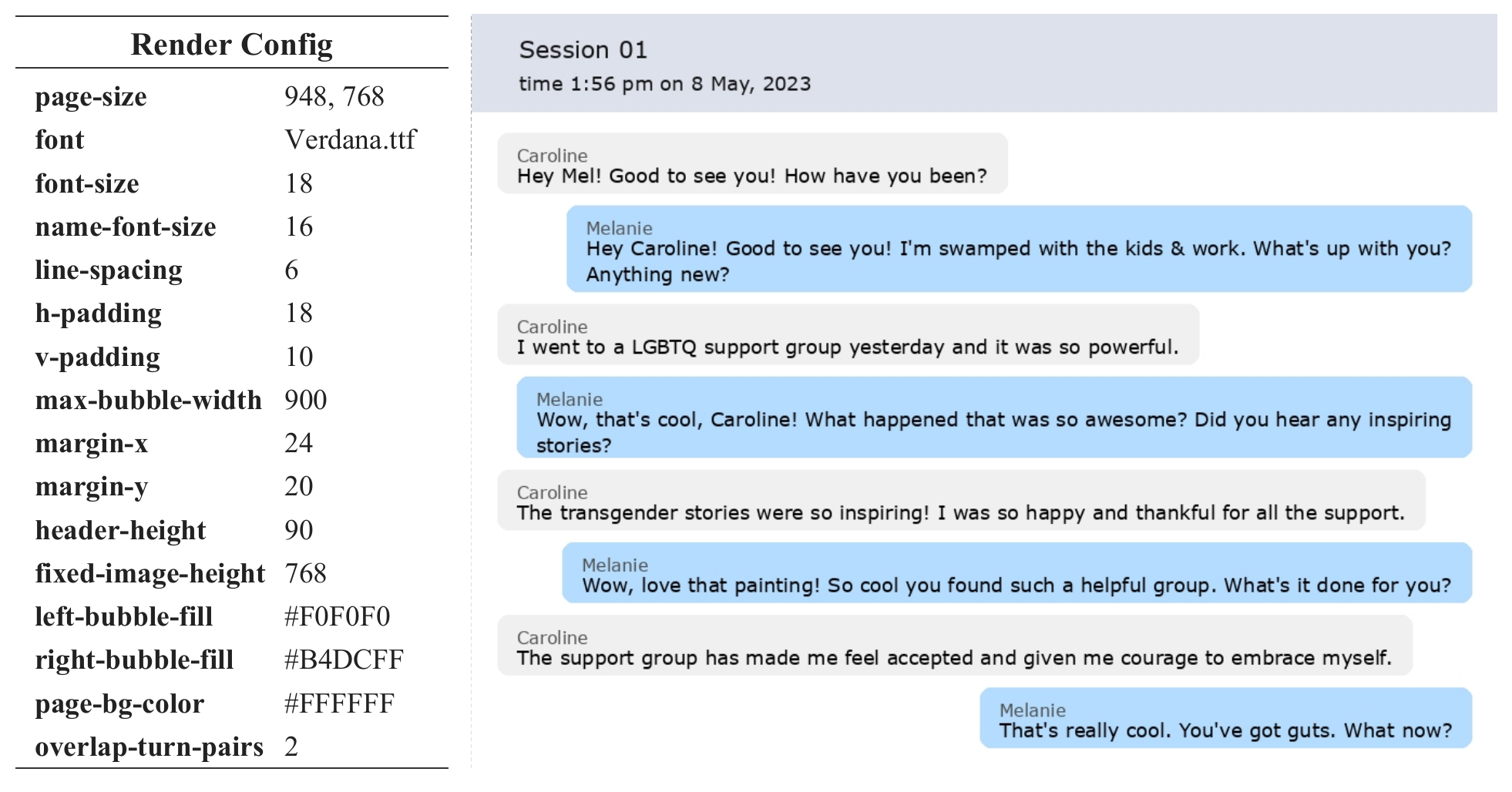}
\caption{Illustration of the Visual Memory Rendering Template Used in \method{}. The left panel summarizes the main rendering parameter configuration, and the right panel shows an example rendered visual memory unit.}
\label{fig:render_config}
\end{figure}

%% file: tables/overall_longmemeval_2b.tex
\begin{table*}[t!]
\centering
\caption{Overall Performance on the LongMemEval Dataset on Qwen3-VL-2B-Instruct Model. The best results are marked in \textbf{bold}, while the \underline{second-best} results are underlined.}
\label{tab:longmemeval_2b}
\resizebox{\linewidth}{!}{
\begin{tabular}{lcccccccccccccc}
\toprule
\multirow{2}{*}{\textbf{Method}} 
& \multicolumn{2}{c}{\textbf{SS-User}} 
& \multicolumn{2}{c}{\textbf{SS-Asst}} 
& \multicolumn{2}{c}{\textbf{SS-pref}} 
& \multicolumn{2}{c}{\textbf{Multi-S}} 
& \multicolumn{2}{c}{\textbf{Know. Upd}} 
& \multicolumn{2}{c}{\textbf{Temp. Reas}} 
& \multicolumn{2}{c}{\textbf{Overall}} \\
\cmidrule(lr){2-3}\cmidrule(lr){4-5}\cmidrule(lr){6-7}\cmidrule(lr){8-9}\cmidrule(lr){10-11}\cmidrule(lr){12-13}\cmidrule(lr){14-15}
& \textbf{Acc} & \textbf{F1} 
& \textbf{Acc} & \textbf{F1} 
& \textbf{Acc} & \textbf{F1} 
& \textbf{Acc} & \textbf{F1} 
& \textbf{Acc} & \textbf{F1} 
& \textbf{Acc} & \textbf{F1} 
& \textbf{Acc} & \textbf{F1} \\
\midrule

\rowcolor{gray!20}\multicolumn{15}{l}{\textbf{Qwen3-VL-2B-Instruct}} \\
\midrule
Text RAG & 77.14 & 37.68 & 80.36 & \underline{59.44} & 23.33 & \underline{12.51} & 22.56 & \underline{15.65} & \underline{55.13} & 27.82 & \underline{27.82} & \underline{20.61} & 43.20 & \underline{26.67} \\
LightMem~\cite{fang2025lightmem} & 55.71 & 13.93 & 16.07 & 6.89 & \textbf{66.67} & 11.80 & 13.53 & 4.78 & 38.46 & 7.74 & 26.32 & 8.15 & 30.20 & 8.08 \\
MemOS~\cite{li2025memos} & \underline{80.00} & 25.17 & \textbf{92.86} & 27.87 & \underline{53.33} & \textbf{14.19} & \textbf{29.32} & 6.22 & \textbf{56.41} & 12.72 & \textbf{33.83} & 8.61 & \textbf{50.40} & 13.55 \\
MemoryOS~\cite{kang2025memory} & 71.43 & \underline{48.21} & 78.57 & 52.59 & 13.33 & 3.39 & \underline{25.56} & 12.58 & 50.00 & 30.79 & 14.29 & 14.99 & 38.00 & 24.98 \\
EverMemOS~\cite{hu2026evermemos} & 54.29 & 12.81 & 32.14 & 14.26 & 23.33 & 11.11 & 21.05 & 7.88 & 25.64 & 7.91 & 18.05 & 10.14 & 27.00 & 10.19 \\
\hdashline
MemOCR~\cite{shi2026memocr} & 21.43 & 18.61 & 23.21 & 22.99 & 13.33 & 9.79 & 9.02 & 8.09 & 37.18 & \underline{31.44} & 18.80 & 19.26 & 19.60 & 17.95 \\
\method{} & \textbf{91.43} & \textbf{75.38} & \underline{91.07} & \textbf{76.03} & 30.00 & 6.50 & 23.31 & \textbf{21.07} & 48.72 & \textbf{37.34} & 26.32 & \textbf{27.65} & \underline{45.60} & \textbf{38.24} \\
\bottomrule
\end{tabular}
}
\end{table*}

%% file: tables/render_height.tex
\begin{table}[t]
\centering
\caption{Effect of Maximum Rendering Height on the LoCoMo Dataset Using Qwen3-VL-8B-Instruct Model.}
\label{tab:rendering_height}
\begin{tabular}{lcccc}
\toprule
\multirow{2}{*}{\textbf{Method}} & \multicolumn{2}{c}{\textbf{Top-5}} & \multicolumn{2}{c}{\textbf{Top-10}} \\
\cmidrule(lr){2-3} \cmidrule(lr){4-5}
 & \textbf{Acc} & \textbf{F1} & \textbf{Acc} & \textbf{F1} \\
\midrule
\method{} (Full Session)      & 70.26 & 52.93             & \underline{74.87}             & 54.11 \\
\method{} (Fixed Length 1024) & \underline{70.78}             & \underline{54.87} & 73.25             & \underline{56.18} \\
\method{} (Fixed Length 768)  & \textbf{72.73}    & \textbf{55.03}    & \textbf{75.13}    & \textbf{56.46} \\
\method{} (Fixed Length 512)  & 70.19             & 54.66             & 72.34             & 54.93 \\
\bottomrule
\end{tabular}
\end{table}

%% file: figures/llm_judge_prompt.tex
\begin{figure*}[t]
    \centering
    \includegraphics[width=0.9\linewidth]{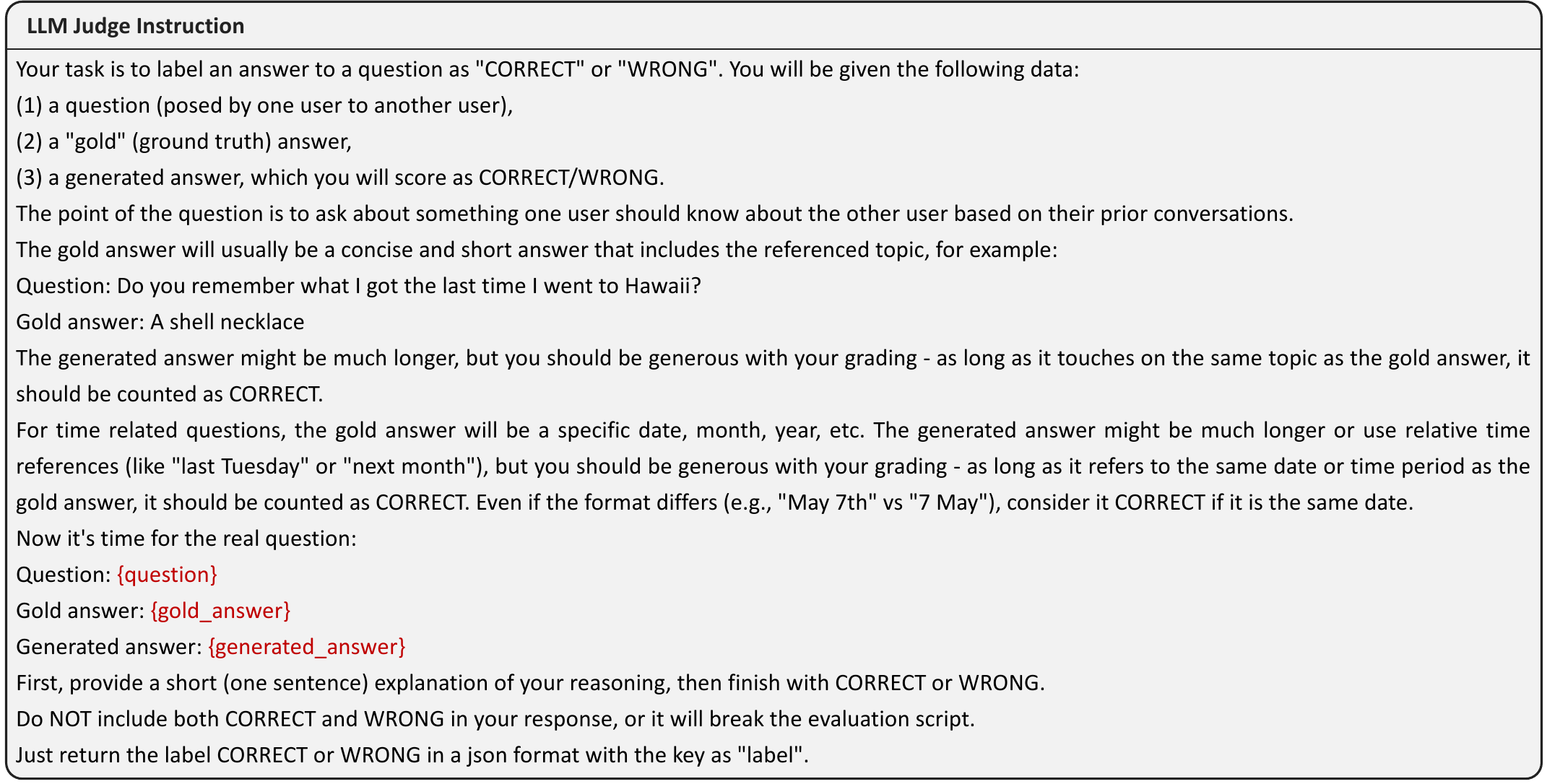}
    \caption{Prompt used for LLM-as-a-Judge evaluation with GLM-5.}
    \label{fig:glm5_judge_prompt}
\end{figure*}

%% file: figures/appendix_case1.tex
\begin{figure*}[t]
\centering
\includegraphics[width=0.8\linewidth]{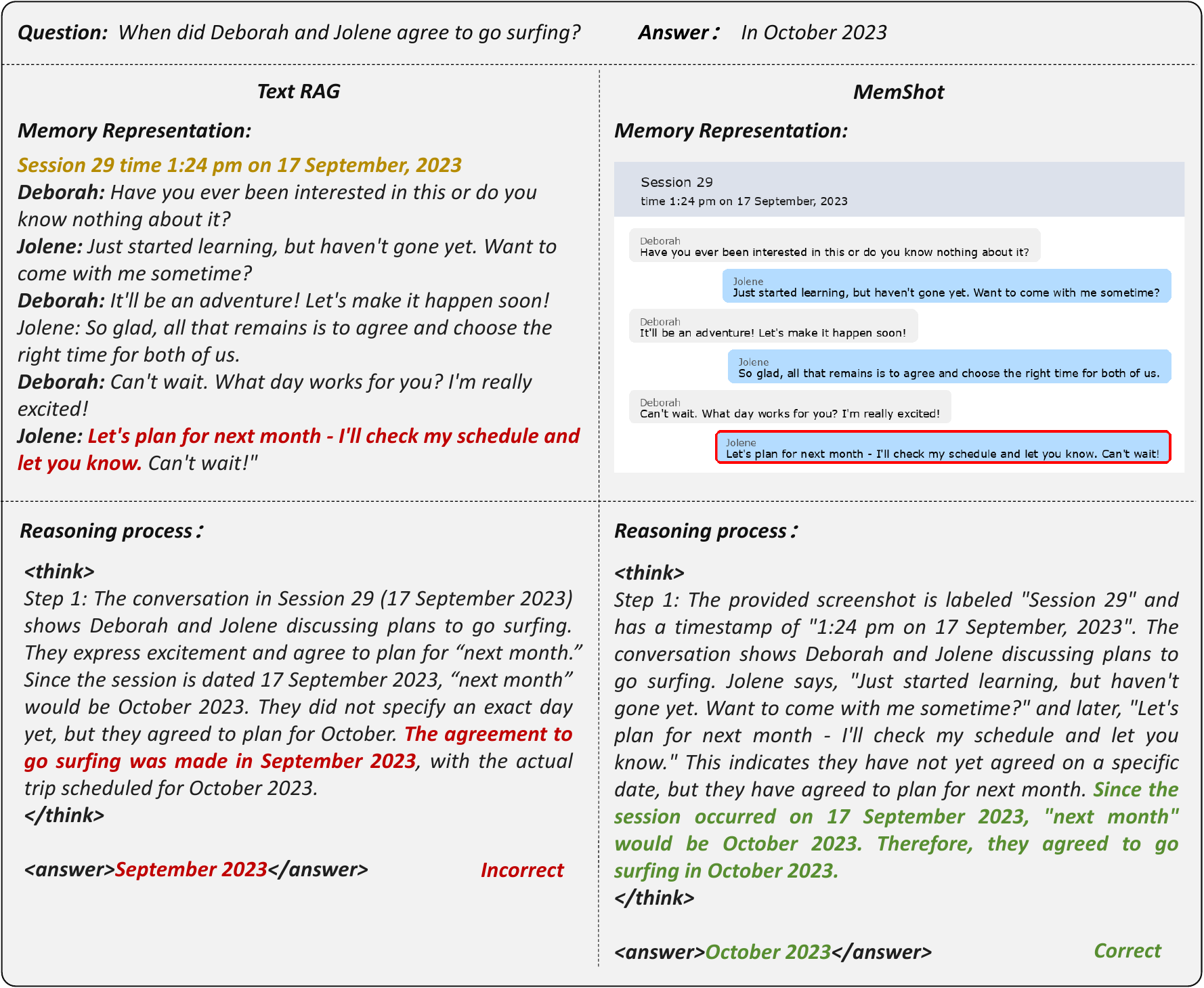}
\caption{Case Study on Temporal Reasoning Scenario with Text RAG and \method{}.}
\label{fig:appendix_case1}
\end{figure*}

%% file: figures/appendix_case2.tex
\begin{figure*}[t]
\centering
\includegraphics[width=0.8\linewidth]{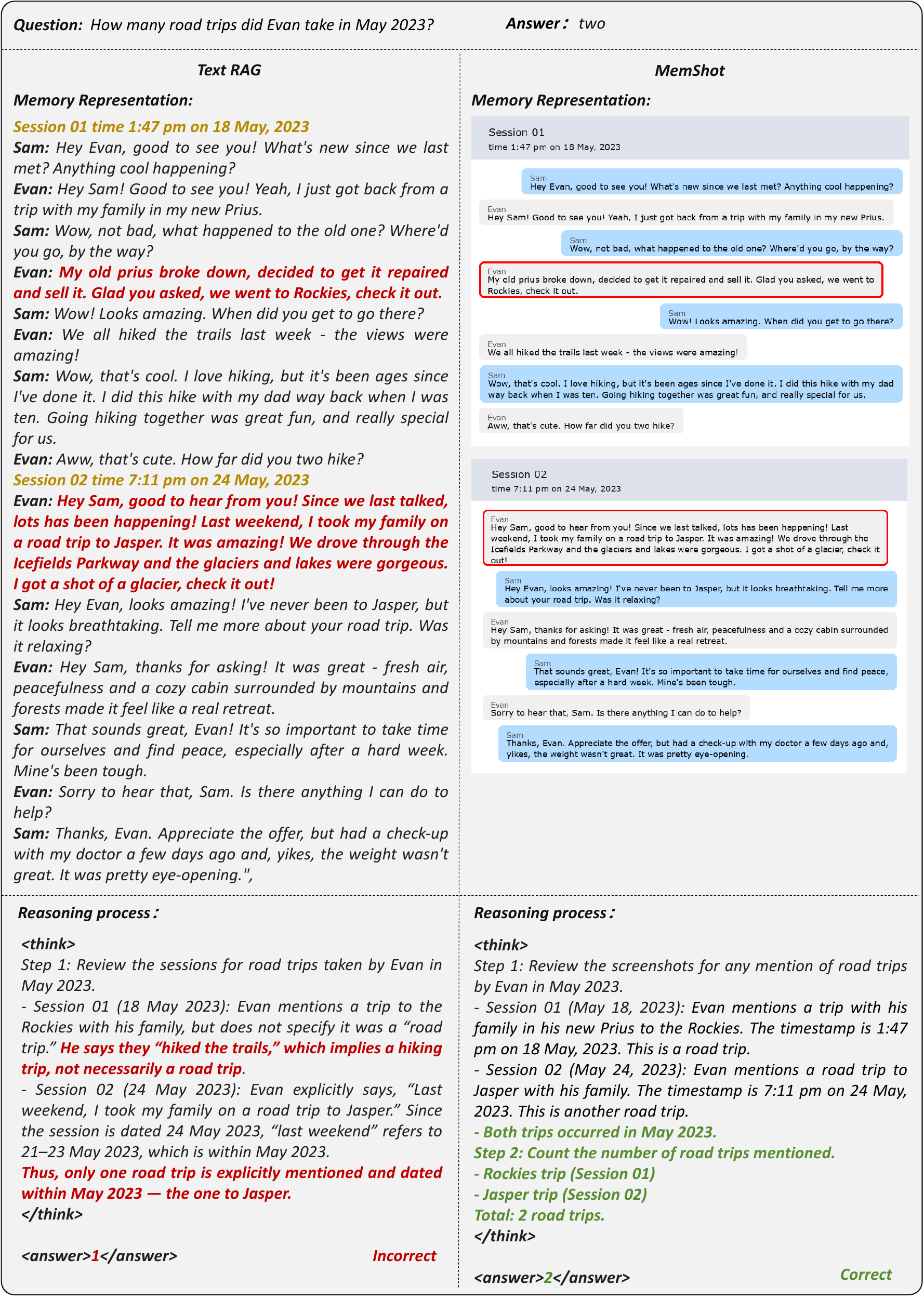}
\caption{Case Study on Multi-Session Evidence Aggregation Scenario with Text RAG and \method{}.}
\label{fig:appendix_case2}
\end{figure*}

%% file: figures/text_rag_prompt.tex
\begin{figure*}[t]
\centering
\includegraphics[width=0.9\linewidth]{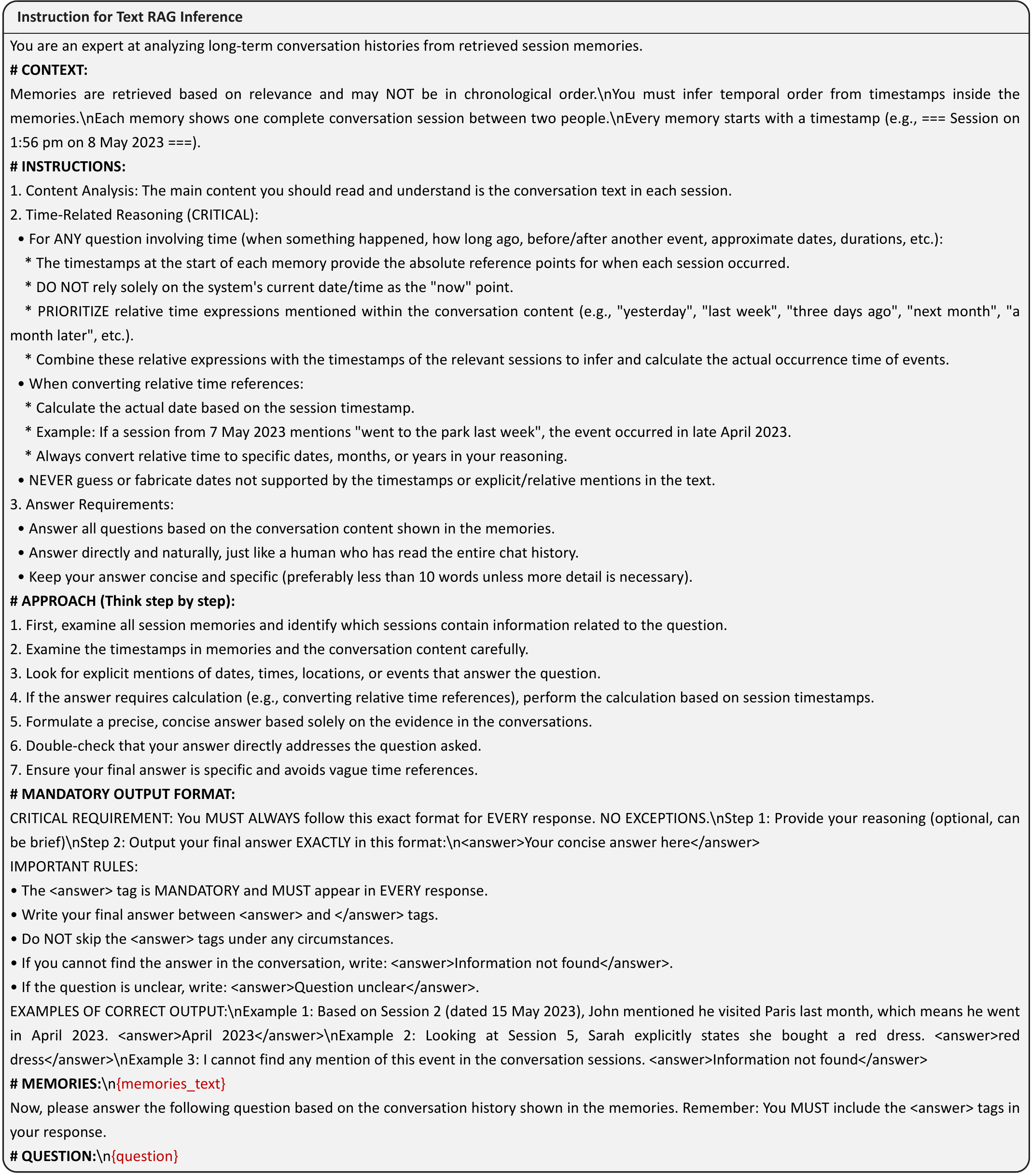}
\caption{Prompt Template for Text RAG Inference Used in Our Experiments.}
\label{fig:text_rag_prompt}
\end{figure*}

%% file: figures/memshot_prompt.tex
\begin{figure*}[t]
\centering
\includegraphics[width=0.9\linewidth]{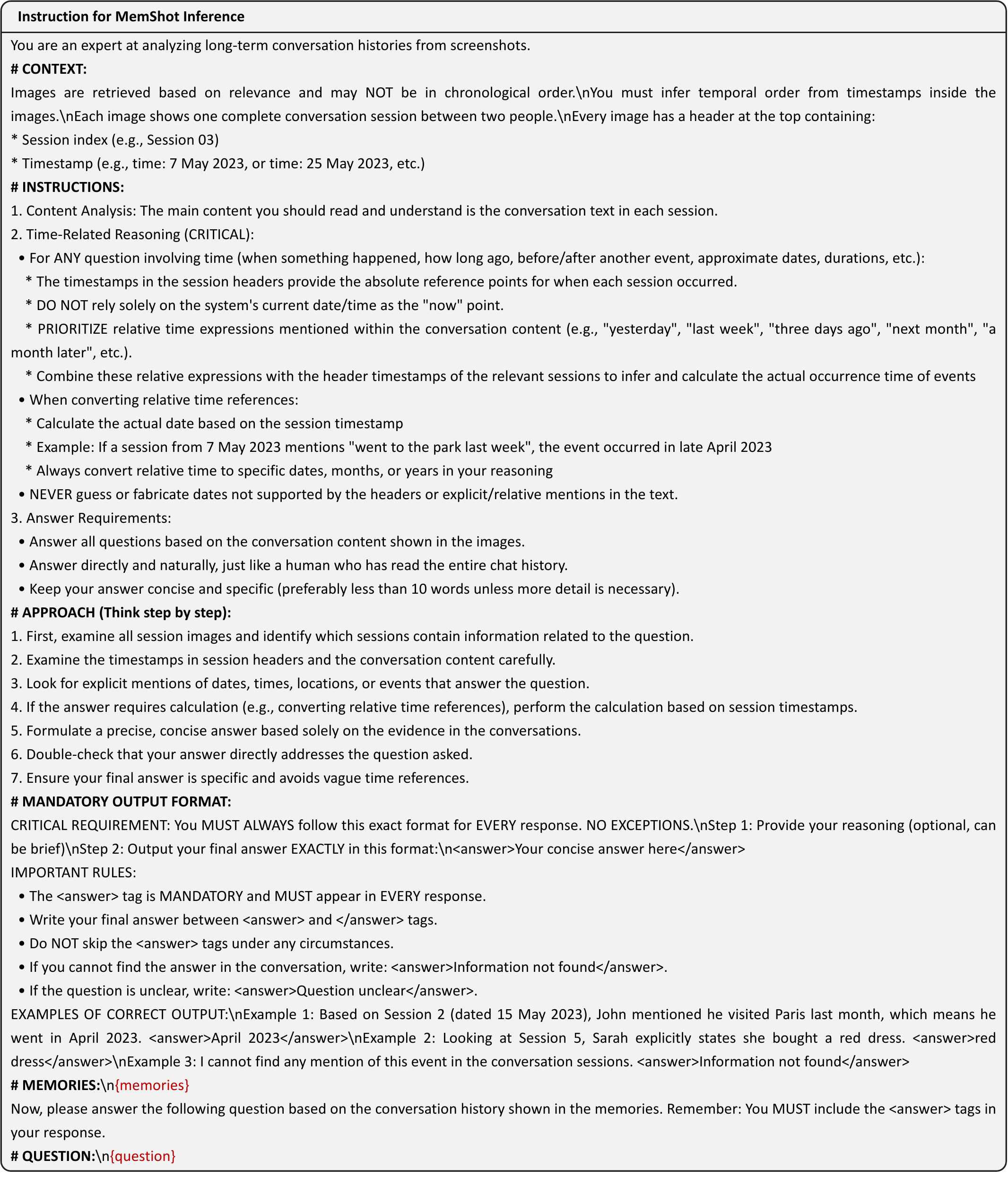}
\caption{Prompt Template for \method{} Inference Used in Our Experiments.}
\label{fig:memshot_prompt}
\end{figure*}

%% file: figures/rubric_cot_analysis_prompt.tex
\begin{figure*}[t]
\centering
\includegraphics[width=0.9\linewidth]{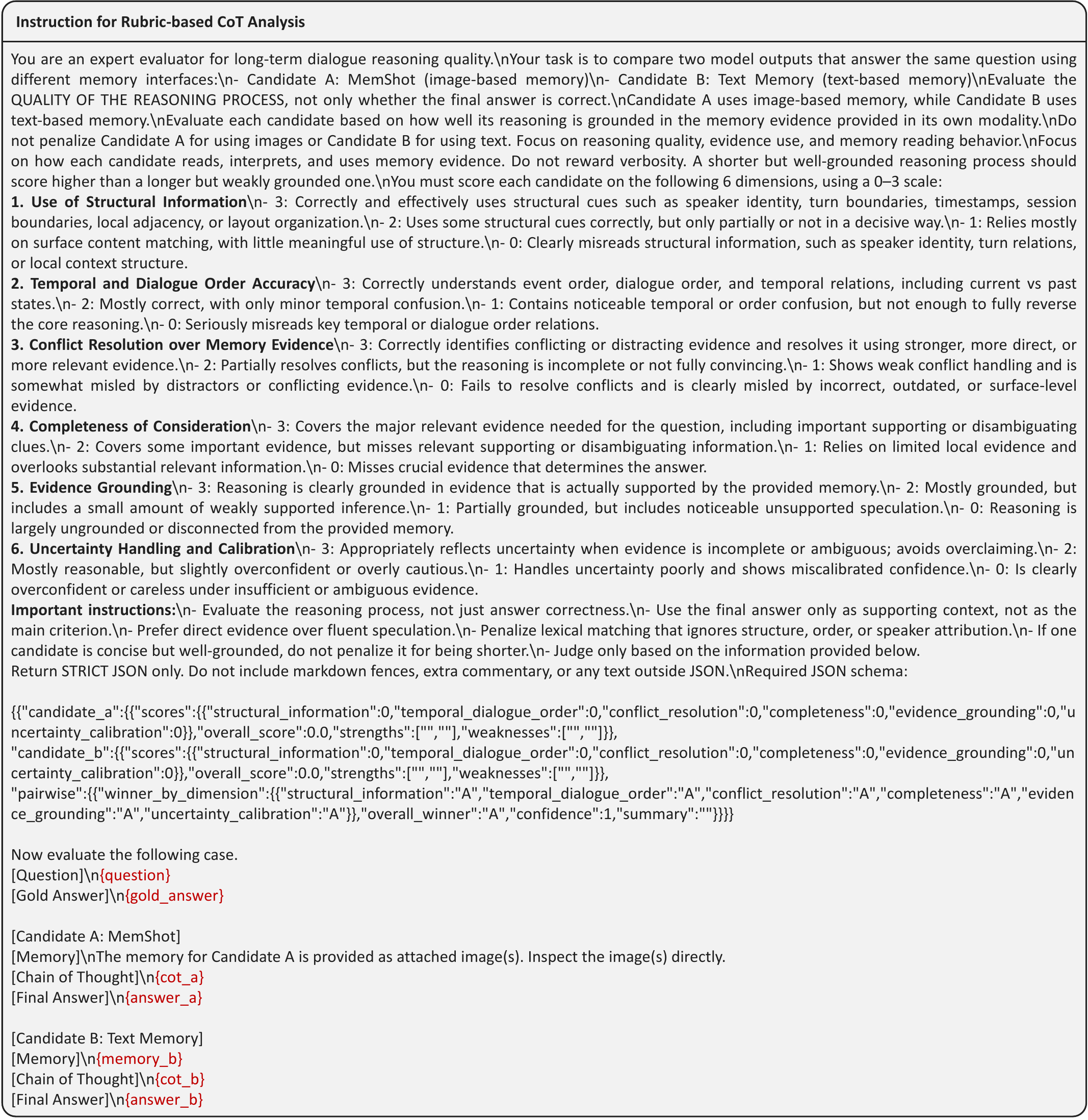}
\caption{Prompt Template for Rubric-Based Chain-of-Thought Analysis Used in Our Experiments.}
\label{fig:rubric_cot_analysis_prompt}
\end{figure*}

%% file: sample-base.bib
@String{Computer = "{IEEE} Computer" }

@article{xin2026metamem,
    author = {Xin, Haidong and Li, Xinze and Liu, Zhenghao and Yan, Yukun and Wang, Shuo and Yang, Cheng and Gu, Yu and Yu, Ge and Sun, Maosong},
    journal = {ArXiv preprint},
    title = {MetaMem: Evolving Meta-Memory for Knowledge Utilization through Self-Reflective Symbolic Optimization},
    url = {https://arxiv.org/abs/2602.11182},
    volume = {abs/2602.11182},
    year = {2026}
}

@article{hu2026evermemos,
    author = {Hu, Chuanrui and Gao, Xingze and Zhou, Zuyi and Xu, Dannong and Bai, Yi and Li, Xintong and Zhang, Hui and Li, Tong and Zhang, Chong and Bing, Lidong and others},
    journal = {ArXiv preprint},
    title = {EverMemOS: A Self-Organizing Memory Operating System for Structured Long-Horizon Reasoning},
    url = {https://arxiv.org/abs/2601.02163},
    volume = {abs/2601.02163},
    year = {2026}
}

@article{liu2026knowledge,
    author = {Liu, Zhenghao and Huang, Pengcheng and Xu, Zhipeng and Li, Xinze and Liu, Shuliang and Peng, Chunyi and Xin, Haidong and Yan, Yukun and Wang, Shuo and Han, Xu and others},
    journal = {AI Open},
    publisher = {Elsevier},
    title = {Knowledge intensive agents},
    year = {2026}
}

@article{li2025memos,
    author = {Li, Zhiyu and Xi, Chenyang and Li, Chunyu and Chen, Ding and Chen, Boyu and Song, Shichao and Niu, Simin and Wang, Hanyu and Yang, Jiawei and Tang, Chen and others},
    journal = {ArXiv preprint},
    title = {Memos: A memory os for ai system},
    url = {https://arxiv.org/abs/2507.03724},
    volume = {abs/2507.03724},
    year = {2025}
}

@article{chhikara2025mem0,
    author = {Chhikara, Prateek and Khant, Dev and Aryan, Saket and Singh, Taranjeet and Yadav, Deshraj},
    journal = {ArXiv preprint},
    title = {Mem0: Building production-ready ai agents with scalable long-term memory},
    url = {https://arxiv.org/abs/2504.19413},
    volume = {abs/2504.19413},
    year = {2025}
}

@inproceedings{kang2025memory,
    author = {Kang, Jiazheng and Ji, Mingming and Zhao, Zhe and Bai, Ting},
    booktitle = {Proceedings of the 2025 Conference on Empirical Methods in Natural Language Processing},
    pages = {25972--25981},
    title = {Memory os of ai agent},
    year = {2025}
}

@article{shi2026memocr,
    author = {Shi, Yaorui and Liu, Shugui and Yang, Yu and Mao, Wenyu and Chen, Yuxin and Gu, Qi and Su, Hui and Cai, Xunliang and Wang, Xiang and Zhang, An},
    journal = {ArXiv preprint},
    title = {MemOCR: Layout-Aware Visual Memory for Efficient Long-Horizon Reasoning},
    url = {https://arxiv.org/abs/2601.21468},
    volume = {abs/2601.21468},
    year = {2026}
}

@article{bai2025qwen3,
    author = {Bai, Shuai and Cai, Yuxuan and Chen, Ruizhe and Chen, Keqin and Chen, Xionghui and Cheng, Zesen and Deng, Lianghao and Ding, Wei and Gao, Chang and Ge, Chunjiang and others},
    journal = {ArXiv preprint},
    title = {Qwen3-vl technical report},
    url = {https://arxiv.org/abs/2511.21631},
    volume = {abs/2511.21631},
    year = {2025}
}

@article{li2026qwen3,
    author = {Li, Mingxin and Zhang, Yanzhao and Long, Dingkun and Chen, Keqin and Song, Sibo and Bai, Shuai and Yang, Zhibo and Xie, Pengjun and Yang, An and Liu, Dayiheng and others},
    journal = {ArXiv preprint},
    title = {Qwen3-VL-Embedding and Qwen3-VL-Reranker: A Unified Framework for State-of-the-Art Multimodal Retrieval and Ranking},
    url = {https://arxiv.org/abs/2601.04720},
    volume = {abs/2601.04720},
    year = {2026}
}

@article{fang2025lightmem,
    author = {Fang, Jizhan and Deng, Xinle and Xu, Haoming and Jiang, Ziyan and Tang, Yuqi and Xu, Ziwen and Deng, Shumin and Yao, Yunzhi and Wang, Mengru and Qiao, Shuofei and others},
    journal = {ArXiv preprint},
    title = {Lightmem: Lightweight and efficient memory-augmented generation},
    url = {https://arxiv.org/abs/2510.18866},
    volume = {abs/2510.18866},
    year = {2025}
}

@inproceedings{maharana2024evaluating,
    author = {Maharana, Adyasha and Lee, Dong-Ho and Tulyakov, Sergey and Bansal, Mohit and Barbieri, Francesco and Fang, Yuwei},
    booktitle = {Proceedings of the 62nd Annual Meeting of the Association for Computational Linguistics (Volume 1: Long Papers)},
    pages = {13851--13870},
    title = {Evaluating very long-term conversational memory of llm agents},
    year = {2024}
}

@article{wu2024longmemeval,
    author = {Wu, Di and Wang, Hongwei and Yu, Wenhao and Zhang, Yuwei and Chang, Kai-Wei and Yu, Dong},
    journal = {ArXiv preprint},
    title = {Longmemeval: Benchmarking chat assistants on long-term interactive memory},
    url = {https://arxiv.org/abs/2410.10813},
    volume = {abs/2410.10813},
    year = {2024}
}

@article{zeng2026glm,
    author = {Zeng, Aohan and Lv, Xin and Hou, Zhenyu and Du, Zhengxiao and Zheng, Qinkai and Chen, Bin and Yin, Da and Ge, Chendi and Xie, Chengxing and Wang, Cunxiang and others},
    journal = {ArXiv preprint},
    title = {GLM-5: from Vibe Coding to Agentic Engineering},
    url = {https://arxiv.org/abs/2602.15763},
    volume = {abs/2602.15763},
    year = {2026}
}

@article{wei2025deepseek,
    author = {Wei, Haoran and Sun, Yaofeng and Li, Yukun},
    journal = {ArXiv preprint},
    title = {Deepseek-ocr: Contexts optical compression},
    url = {https://arxiv.org/abs/2510.18234},
    volume = {abs/2510.18234},
    year = {2025}
}

@article{cheng2025glyph,
    author = {Cheng, Jiale and Liu, Yusen and Zhang, Xinyu and Fei, Yulin and Hong, Wenyi and Lyu, Ruiliang and Wang, Weihan and Su, Zhe and Gu, Xiaotao and Liu, Xiao and others},
    journal = {ArXiv preprint},
    title = {Glyph: Scaling context windows via visual-text compression},
    url = {https://arxiv.org/abs/2510.17800},
    volume = {abs/2510.17800},
    year = {2025}
}

@techreport{seed2025seed2,
    author = {Seed, ByteDance},
    institution = {Technical report (model card), February 2026. URL https://lf3-static~…},
    title = {Seed 2.0 Model Card: Towards Intelligence Frontier for Real-World Complexity},
    year = {2026}
}

@article{yang2025qwen3,
    author = {Yang, An and Li, Anfeng and Yang, Baosong and Zhang, Beichen and Hui, Binyuan and Zheng, Bo and Yu, Bowen and Gao, Chang and Huang, Chengen and Lv, Chenxu and others},
    journal = {ArXiv preprint},
    title = {Qwen3 technical report},
    url = {https://arxiv.org/abs/2505.09388},
    volume = {abs/2505.09388},
    year = {2025}
}

@article{liu2025comprehensive,
    author = {Liu, Jiaheng and Zhu, Dawei and Bai, Zhiqi and He, Yancheng and Liao, Huanxuan and Que, Haoran and Wang, Zekun and Zhang, Chenchen and Zhang, Ge and Zhang, Jiebin and others},
    journal = {ArXiv preprint},
    title = {A comprehensive survey on long context language modeling},
    url = {https://arxiv.org/abs/2503.17407},
    volume = {abs/2503.17407},
    year = {2025}
}

@inproceedings{wang2024beyond,
    author = {Xindi Wang and
Mahsa Salmani and
Parsa Omidi and
Xiangyu Ren and
Mehdi Rezagholizadeh and
Armaghan Eshaghi},
    bibsource = {dblp computer science bibliography, https://dblp.org},
    booktitle = {Proceedings of the Thirty-Third International Joint Conference on
Artificial Intelligence, {IJCAI} 2024, Jeju, South Korea, August 3-9,
2024},
    pages = {8299--8307},
    publisher = {ijcai.org},
    title = {Beyond the Limits: {A} Survey of Techniques to Extend the Context
Length in Large Language Models},
    url = {https://www.ijcai.org/proceedings/2024/917},
    year = {2024}
}

@article{mao2025meta,
    author = {Mao, Yufan and Ye, Hanjing and Dong, Wenlong and Zhang, Chengjie and Zhang, Hong},
    journal = {ArXiv preprint},
    title = {Meta-Memory: Retrieving and Integrating Semantic-Spatial Memories for Robot Spatial Reasoning},
    url = {https://arxiv.org/abs/2509.20754},
    volume = {abs/2509.20754},
    year = {2025}
}

@article{hassabis2007deconstructing,
    author = {Hassabis, Demis and Maguire, Eleanor A},
    journal = {Trends in cognitive sciences},
    number = {7},
    pages = {299--306},
    publisher = {Elsevier},
    title = {Deconstructing episodic memory with construction},
    volume = {11},
    year = {2007}
}

@article{nolden2024prediction,
    author = {Nolden, Sophie and Turan, G{\"o}zem and G{\"u}ler, Berna and G{\"u}nseli, Eren},
    journal = {Neuroscience \& Biobehavioral Reviews},
    pages = {105533},
    publisher = {Elsevier},
    title = {Prediction error and event segmentation in episodic memory},
    volume = {157},
    year = {2024}
}

@article{laing2025event,
    author = {Laing, Patrick AF and Dunsmoor, Joseph E},
    journal = {Journal of Cognitive Neuroscience},
    number = {1},
    pages = {110--134},
    publisher = {MIT Press 255 Main Street, 9th Floor, Cambridge, Massachusetts 02142, USA~…},
    title = {Event segmentation promotes the reorganization of emotional memory},
    volume = {37},
    year = {2025}
}

@article{fan2025improving,
    author = {Fan, Sinan and Xie, Liang and Shen, Chen and Teng, Ge and Yuan, Xiaosong and Zhang, Xiaofeng and Huang, Chenxi and Wang, Wenxiao and He, Xiaofei and Ye, Jieping},
    journal = {ArXiv preprint},
    title = {Improving complex reasoning with dynamic prompt corruption: A soft prompt optimization approach},
    url = {https://arxiv.org/abs/2503.13208},
    volume = {abs/2503.13208},
    year = {2025}
}

@article{liu2024lost,
    address = {Cambridge, MA},
    author = {Liu, Nelson F.  and
Lin, Kevin  and
Hewitt, John  and
Paranjape, Ashwin  and
Bevilacqua, Michele  and
Petroni, Fabio  and
Liang, Percy},
    doi = {10.1162/tacl_a_00638},
    journal = {Transactions of the Association for Computational Linguistics},
    pages = {157--173},
    publisher = {MIT Press},
    title = {Lost in the Middle: How Language Models Use Long Contexts},
    url = {https://aclanthology.org/2024.tacl-1.9},
    volume = {12},
    year = {2024}
}

@inproceedings{zhou2025llm,
    author = {Zhou, Zihan and Li, Chong and Chen, Xinyi and Wang, Shuo and Chao, Yu and Li, Zhili and Wang, Haoyu and Shi, Qi and Tan, Zhixing and Han, Xu and others},
    booktitle = {Proceedings of the 63rd Annual Meeting of the Association for Computational Linguistics (Volume 1: Long Papers)},
    pages = {27664--27678},
    title = {LLM$\times$ MapReduce: Simplified Long-Sequence Processing using Large Language Models},
    year = {2025}
}

@inproceedings{lewis2020retrieval,
    author = {Patrick S. H. Lewis and
Ethan Perez and
Aleksandra Piktus and
Fabio Petroni and
Vladimir Karpukhin and
Naman Goyal and
Heinrich K{\"{u}}ttler and
Mike Lewis and
Wen{-}tau Yih and
Tim Rockt{\"{a}}schel and
Sebastian Riedel and
Douwe Kiela},
    bibsource = {dblp computer science bibliography, https://dblp.org},
    booktitle = {Advances in Neural Information Processing Systems 33: Annual Conference
on Neural Information Processing Systems 2020, NeurIPS 2020, December
6-12, 2020, virtual},
    editor = {Hugo Larochelle and
Marc'Aurelio Ranzato and
Raia Hadsell and
Maria{-}Florina Balcan and
Hsuan{-}Tien Lin},
    title = {Retrieval-Augmented Generation for Knowledge-Intensive {NLP} Tasks},
    url = {https://proceedings.neurips.cc/paper/2020/hash/6b493230205f780e1bc26945df7481e5-Abstract.html},
    year = {2020}
}

@article{zeidman2015constructing,
    author = {Zeidman, Peter and Mullally, Sin{\'e}ad L and Maguire, Eleanor A},
    journal = {Cerebral Cortex},
    number = {10},
    pages = {3836--3855},
    publisher = {Oxford University Press},
    title = {Constructing, perceiving, and maintaining scenes: hippocampal activity and connectivity},
    volume = {25},
    year = {2015}
}

@inproceedings{wei2022chain,
    author = {Jason Wei and
Xuezhi Wang and
Dale Schuurmans and
Maarten Bosma and
Brian Ichter and
Fei Xia and
Ed H. Chi and
Quoc V. Le and
Denny Zhou},
    bibsource = {dblp computer science bibliography, https://dblp.org},
    booktitle = {Advances in Neural Information Processing Systems 35: Annual Conference
on Neural Information Processing Systems 2022, NeurIPS 2022, New Orleans,
LA, USA, November 28 - December 9, 2022},
    editor = {Sanmi Koyejo and
S. Mohamed and
A. Agarwal and
Danielle Belgrave and
K. Cho and
A. Oh},
    title = {Chain-of-Thought Prompting Elicits Reasoning in Large Language Models},
    url = {http://papers.nips.cc/paper\_files/paper/2022/hash/9d5609613524ecf4f15af0f7b31abca4-Abstract-Conference.html},
    year = {2022}
}

@inproceedings{hashemi2024llm,
    author = {Hashemi, Helia and Eisner, Jason and Rosset, Corby and Van Durme, Benjamin and Kedzie, Chris},
    booktitle = {Proceedings of the 62nd Annual Meeting of the Association for Computational Linguistics (Volume 1: Long Papers)},
    pages = {13806--13834},
    title = {Llm-rubric: A multidimensional, calibrated approach to automated evaluation of natural language texts},
    year = {2024}
}

@inproceedings{gunther2025jina,
    author = {G{\"u}nther, Michael and Sturua, Saba and Akram, Mohammad Kalim and Mohr, Isabelle and Ungureanu, Andrei and Wang, Bo and Eslami, Sedigheh and Martens, Scott and Werk, Maximilian and Wang, Nan and others},
    booktitle = {Proceedings of the 5th Workshop on Multilingual Representation Learning (MRL 2025)},
    pages = {531--550},
    title = {jina-embeddings-v4: Universal embeddings for multimodal multilingual retrieval},
    year = {2025}
}

@article{yu2025minicpm,
    author = {Yu, Tianyu and Wang, Zefan and Wang, Chongyi and Huang, Fuwei and Ma, Wenshuo and He, Zhihui and Cai, Tianchi and Chen, Weize and Huang, Yuxiang and Zhao, Yuanqian and others},
    journal = {ArXiv preprint},
    title = {Minicpm-v 4.5: Cooking efficient mllms via architecture, data, and training recipe},
    url = {https://arxiv.org/abs/2509.18154},
    volume = {abs/2509.18154},
    year = {2025}
}

@article{wang2025internvl3,
    author = {Wang, Weiyun and Gao, Zhangwei and Gu, Lixin and Pu, Hengjun and Cui, Long and Wei, Xingguang and Liu, Zhaoyang and Jing, Linglin and Ye, Shenglong and Shao, Jie and others},
    journal = {ArXiv preprint},
    title = {Internvl3. 5: Advancing open-source multimodal models in versatility, reasoning, and efficiency},
    url = {https://arxiv.org/abs/2508.18265},
    volume = {abs/2508.18265},
    year = {2025}
}

@article{ram2023context,
    address = {Cambridge, MA},
    author = {Ram, Ori  and
Levine, Yoav  and
Dalmedigos, Itay  and
Muhlgay, Dor  and
Shashua, Amnon  and
Leyton-Brown, Kevin  and
Shoham, Yoav},
    doi = {10.1162/tacl_a_00605},
    journal = {Transactions of the Association for Computational Linguistics},
    pages = {1316--1331},
    publisher = {MIT Press},
    title = {In-Context Retrieval-Augmented Language Models},
    url = {https://aclanthology.org/2023.tacl-1.75},
    volume = {11},
    year = {2023}
}

@article{rasmussen2025zep,
    author = {Rasmussen, Preston and Paliychuk, Pavlo and Beauvais, Travis and Ryan, Jack and Chalef, Daniel},
    journal = {ArXiv preprint},
    title = {Zep: a temporal knowledge graph architecture for agent memory},
    url = {https://arxiv.org/abs/2501.13956},
    volume = {abs/2501.13956},
    year = {2025}
}

@article{xu2025mem,
    author = {Xu, Wujiang and Liang, Zujie and Mei, Kai and Gao, Hang and Tan, Juntao and Zhang, Yongfeng},
    journal = {ArXiv preprint},
    title = {A-mem: Agentic memory for llm agents},
    url = {https://arxiv.org/abs/2502.12110},
    volume = {abs/2502.12110},
    year = {2025}
}

@article{wu2025sgmem,
    author = {Wu, Yaxiong and Zhang, Yongyue and Liang, Sheng and Liu, Yong},
    journal = {ArXiv preprint},
    title = {Sgmem: Sentence graph memory for long-term conversational agents},
    url = {https://arxiv.org/abs/2509.21212},
    volume = {abs/2509.21212},
    year = {2025}
}

@inproceedings{tan2025prospect,
    author = {Tan, Zhen and Yan, Jun and Hsu, I-Hung and Han, Rujun and Wang, Zifeng and Le, Long and Song, Yiwen and Chen, Yanfei and Palangi, Hamid and Lee, George and others},
    booktitle = {Proceedings of the 63rd Annual Meeting of the Association for Computational Linguistics (Volume 1: Long Papers)},
    pages = {8416--8439},
    title = {In prospect and retrospect: Reflective memory management for long-term personalized dialogue agents},
    year = {2025}
}

@article{du2025rethinking,
    author = {Du, Yiming and Huang, Wenyu and Zheng, Danna and Wang, Zhaowei and Montella, Sebastien and Lapata, Mirella and Wong, Kam-Fai and Pan, Jeff Z},
    journal = {ArXiv preprint},
    title = {Rethinking Memory in LLM based Agents: Representations, Operations, and Emerging Topics},
    url = {https://arxiv.org/abs/2505.00675},
    volume = {abs/2505.00675},
    year = {2025}
}

@article{hu2025memory,
    author = {Hu, Yuyang and Liu, Shichun and Yue, Yanwei and Zhang, Guibin and Liu, Boyang and Zhu, Fangyi and Lin, Jiahang and Guo, Honglin and Dou, Shihan and Xi, Zhiheng and others},
    journal = {ArXiv preprint},
    title = {Memory in the age of ai agents},
    url = {https://arxiv.org/abs/2512.13564},
    volume = {abs/2512.13564},
    year = {2025}
}

@article{jiang2026anatomy,
    author = {Jiang, Dongming and Li, Yi and Wei, Songtao and Yang, Jinxin and Kishore, Ayushi and Zhao, Alysa and Kang, Dingyi and Hu, Xu and Chen, Feng and Li, Qiannan and others},
    journal = {ArXiv preprint},
    title = {Anatomy of Agentic Memory: Taxonomy and Empirical Analysis of Evaluation and System Limitations},
    url = {https://arxiv.org/abs/2602.19320},
    volume = {abs/2602.19320},
    year = {2026}
}

@inproceedings{wang2024leveraging,
    author = {Alex Jinpeng Wang and
Linjie Li and
Yiqi Lin and
Min Li and
Lijuan Wang and
Mike Zheng Shou},
    bibsource = {dblp computer science bibliography, https://dblp.org},
    booktitle = {Advances in Neural Information Processing Systems 38: Annual Conference
on Neural Information Processing Systems 2024, NeurIPS 2024, Vancouver,
BC, Canada, December 10 - 15, 2024},
    editor = {Amir Globersons and
Lester Mackey and
Danielle Belgrave and
Angela Fan and
Ulrich Paquet and
Jakub M. Tomczak and
Cheng Zhang},
    title = {Leveraging Visual Tokens for Extended Text Contexts in Multi-Modal
Learning},
    url = {http://papers.nips.cc/paper\_files/paper/2024/hash/19f10adb6749b0c9f1ff7610bd01d44d-Abstract-Conference.html},
    year = {2024}
}

@article{lu2024text,
    author = {Lu, Yujie and Li, Xiujun and Fu, Tsu-Jui and Eckstein, Miguel and Wang, William Yang},
    journal = {ArXiv preprint},
    title = {From text to pixel: Advancing long-context understanding in mllms},
    url = {https://arxiv.org/abs/2405.14213},
    volume = {abs/2405.14213},
    year = {2024}
}

@inproceedings{li2025text,
    author = {Li, Yanhong and Lan, Zixuan and Zhou, Jiawei},
    booktitle = {Findings of the Association for Computational Linguistics: EMNLP 2025},
    pages = {10564--10578},
    title = {Text or Pixels? Evaluating Efficiency and Understanding of LLMs with Visual Text Inputs},
    year = {2025}
}

@inproceedings{zhong2024memorybank,
    author = {Wanjun Zhong and
Lianghong Guo and
Qiqi Gao and
He Ye and
Yanlin Wang},
    bibsource = {dblp computer science bibliography, https://dblp.org},
    booktitle = {Thirty-Eighth {AAAI} Conference on Artificial Intelligence, {AAAI}
2024, Thirty-Sixth Conference on Innovative Applications of Artificial
Intelligence, {IAAI} 2024, Fourteenth Symposium on Educational Advances
in Artificial Intelligence, {EAAI} 2014, February 20-27, 2024, Vancouver,
Canada},
    doi = {10.1609/AAAI.V38I17.29946},
    editor = {Michael J. Wooldridge and
Jennifer G. Dy and
Sriraam Natarajan},
    pages = {19724--19731},
    publisher = {{AAAI} Press},
    title = {MemoryBank: Enhancing Large Language Models with Long-Term Memory},
    url = {https://doi.org/10.1609/aaai.v38i17.29946},
    year = {2024}
}

@article{packer2023memgpt,
    author = {Packer, Charles and Fang, Vivian and Patil, Shishir\_G and Lin, Kevin and Wooders, Sarah and Gonzalez, Joseph\_E},
    publisher = {ArXiv},
    title = {MemGPT: towards LLMs as operating systems.},
    year = {2023}
}

@article{bhat2025rethinking,
    author = {Bhat, Sinchana Ramakanth and Rudat, Max and Spiekermann, Jannis and Flores-Herr, Nicolas},
    journal = {ArXiv preprint},
    title = {Rethinking chunk size for long-document retrieval: A multi-dataset analysis},
    url = {https://arxiv.org/abs/2505.21700},
    volume = {abs/2505.21700},
    year = {2025}
}
